\documentclass[11pt,english]{article}
\usepackage{lmodern}
\usepackage[T1]{fontenc}
\usepackage[latin9]{inputenc}
\usepackage{geometry}
\usepackage{color}
\usepackage{babel}
\usepackage{float}
\usepackage{amsmath}
\usepackage{amsthm}
\usepackage{amssymb}
\usepackage{graphicx}
\usepackage{setspace}
\usepackage{natbib}
\bibpunct{(}{)}{;}{a}{}{,}
\usepackage[unicode=true,
 bookmarks=true,bookmarksnumbered=false,bookmarksopen=true,bookmarksopenlevel=2,
 breaklinks=false,pdfborder={0 0 0},backref=false,colorlinks=true]
 {hyperref}
\hypersetup{pdftitle={Robust Variable and Interaction Selection for  
 Logistic Regression and Multiple Index Models},
 pdfauthor={Yang Li},
 linkcolor=blue, citecolor=blue, urlcolor=blue}

\makeatletter

\providecommand{\tabularnewline}{\\}

\theoremstyle{plain}
\newtheorem{thm}{\protect\theoremname}
  \theoremstyle{definition}
  \newtheorem{defn}{\protect\definitionname}

\usepackage{times}
\usepackage{indentfirst}

\usepackage{textpos}
\usepackage{changepage}

\makeatother

  \providecommand{\definitionname}{Definition}
\providecommand{\theoremname}{Theorem}

\begin{document}

\title{Robust Variable and Interaction Selection for  
 Logistic Regression and Multiple Index Models}

\author{Yang Li,$\,$ Jun S. Liu\\
Department of Statistics, Harvard University}
\maketitle
\begin{abstract}
We propose Stepwise cOnditional likelihood variable selection for Discriminant Analysis (SODA) to detect both main and quadratic interaction effects in logistic regression and quadratic discriminant analysis (QDA) models. In the forward stage, SODA adds in important predictors evaluated based on their overall contributions, whereas in the backward stage SODA removes unimportant terms so as to optimize
the extended Bayesian Information Criterion (EBIC).  Compared with existing methods
on QDA variable selections,
SODA can deal with high-dimensional data with the number of predictors much larger
than the sample size and does not require the joint normality assumption
on predictors, leading to much enhanced robustness. 
We further extend SODA to conduct variable selection and model fitting for  multiple index models.
Compared with existing variable selection methods based on the Sliced Inverse Regression (SIR) \citep{li1991sliced}, SODA requires neither the linearity nor the constant variance condition and is much more robust. Our theoretical analyses establish the variable-selection consistency of SODA under high-dimensional settings, and our simulation studies as well as real-data applications demonstrate superior performances of SODA 
in dealing with non-Gaussian  design matrices in both classification problems and multiple index models. 
\end{abstract}

\section{Introduction}

Classification, also known as \textquotedbl{}supervised learning\textquotedbl{},
is a fundamental building block of statistical machine learning.
Applications of statistical classification methods include, for example, cancer diagnosis \citep{tibshirani2002diagnosis}, text
categorization \citep{joachims1998text}, computer vision \citep{phillips1998support}, protein interaction predictions
\citep{chowdhary2009bayesian}, etc. Well-known classification methods include logistic regression, naive Bayes classifier, K-nearest-neighbors, support vector machines  \citep{boser1992training}, and random forests
\citep{breiman2001random}. As important players in this field, linear and quadratic discriminant analysis (LDA and QDA) \citep{anderson1958introduction}
are widely used.  Compared with LDA, QDA is able to exploit interaction effects of predictors.

With rapid technical advances in data collection, it has become
common that the number of predictors is much larger than the number
of observations, which is also known as the ``large $p$ small $n$''
problem. For example, in gene expression microarray analysis, usually
$n$ is in hundreds of samples, whereas $p$ is in thousands of genes
\citep{efron2010large}. In a typical genome-wide association study,
$n$ is in the order of a few thousands of subjects, and $p$ is from
several thousands to  millions of SNP markers \citep{waldmann2013evaluation}.
Vanilla LDA or QDA are infeasible when $p>n$ since the sample covariance
matrices are consequently singular. Even in low-dimensional scenarios,
including many irrelevant predictors can significantly impair the classification accuracy.

A number of variable selection methods have been developed for high-dimensional classification problems, of which many focused on imposing regularizations on the LDA model.
For example, \cite{witten2011penalized} proposed to use fused Lasso to penalize  discriminant vectors in Fisher's discriminant
problem. \cite{cai2011direct} proposed to estimate the product of
precision matrix and the difference between two mean vectors directly
through a constrained $L_1$ minimization. \cite{han2013coda} relaxed
the normal assumption of LDA to entertain Gaussian Copula models.
More  developments on high-dimensional LDA can be found in
\cite{guo2007regularized}, \cite{fan2008high}, \cite{clemmensen2011sparse},
\cite{shao2011sparse}, \cite{mai2012direct} and \cite{fan2013optimal}. 

Aforementioned methods work for LDA models with only linear main effects.
In many applications, however, interaction effects may be significant
and scientifically interesting. 
On the other hand,  in moderate to high dimensional situations, including in the model too many noise variables  and their interaction terms can lead to an over-fitting problem more severe than that of  linear discriminant models, resulting in a much impaired prediction accuracy. 
 In recent years, there has been a significant
surge of interest in detecting interaction effects for regression
or classification problems  \citep{simon2012permutation,bien2013lasso,jiang2014variable,fan2015innovated},
which both improves the classification accuracy and is of scientific interest.
In this article, we use the term ``interaction'' to refer to all second-order effects, including both two-way interactions $X_{i}X_{j}$ with $i\neq j$ and quadratic
terms  $X_{i}^{2}$. 

\begin{figure}[h]
\noindent \begin{centering}
\hspace{-5pt}\includegraphics[viewport=0 15 324 310,scale=0.68]{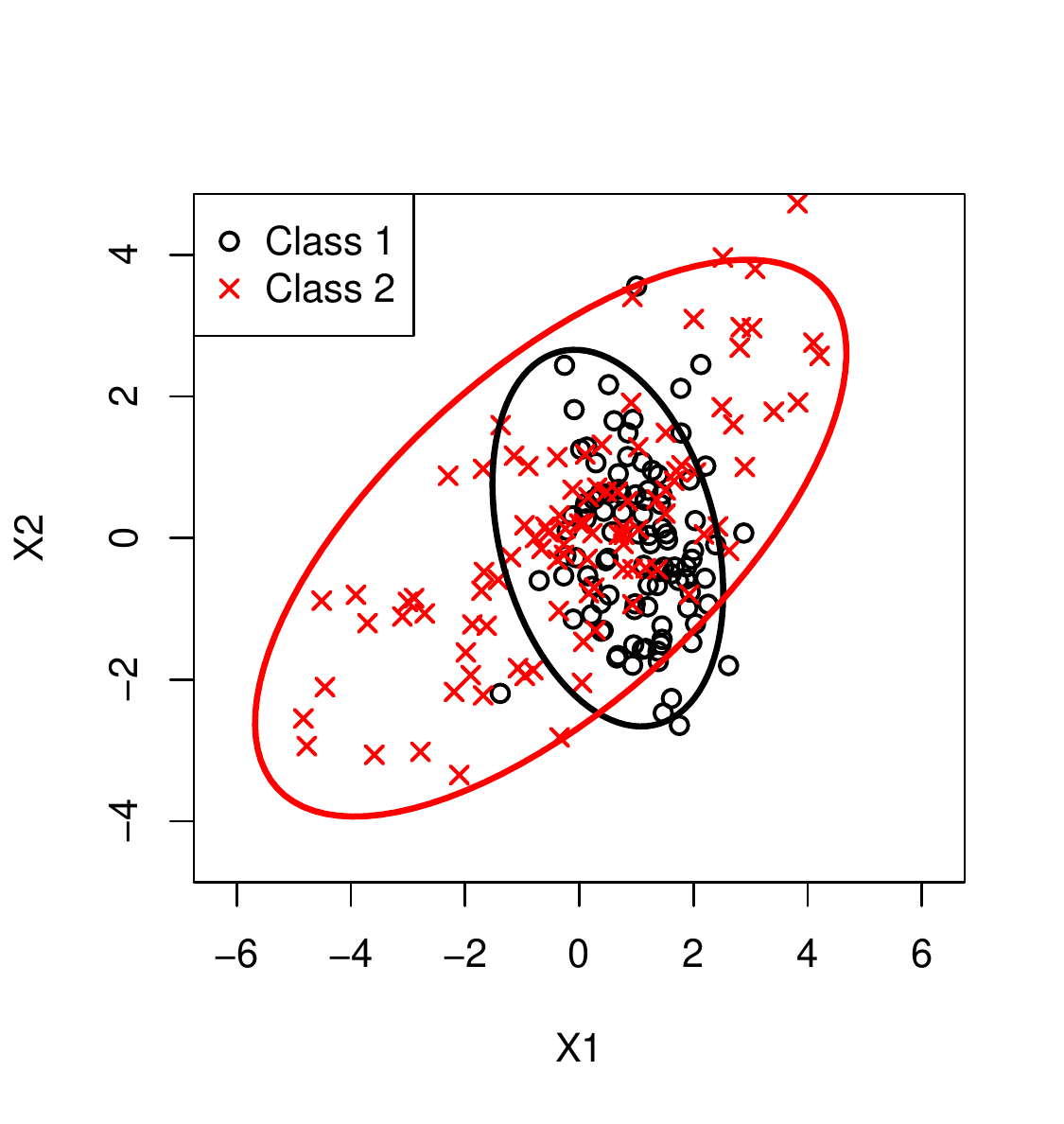}$\;\;\;$\includegraphics[viewport=0bp 15bp 324bp 310bp,scale=0.68]{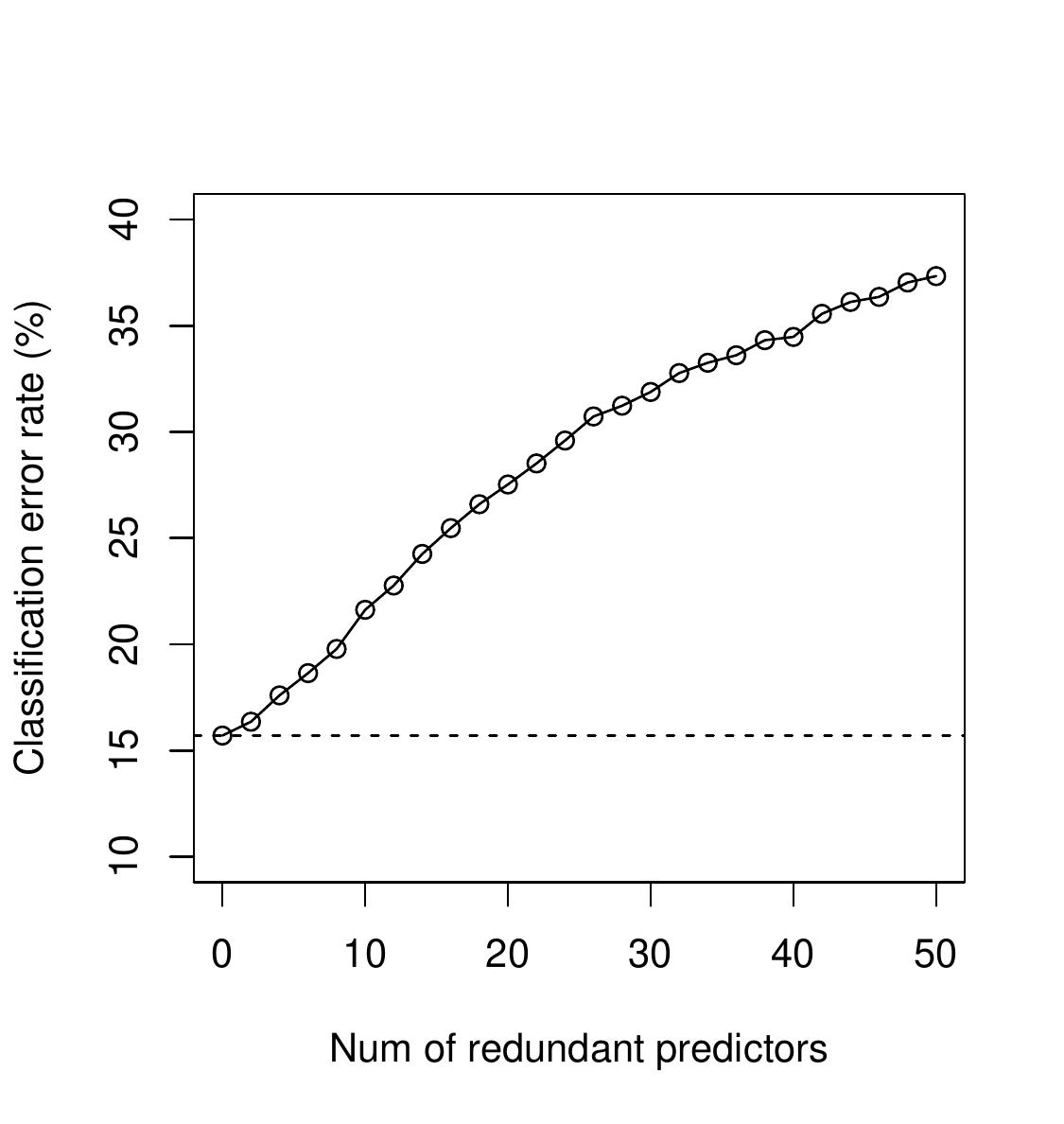}
\par\end{centering}

\caption{A two-class Gaussian classification problem, where Class 1 samples were drawn from $N(\boldsymbol{\mu}_{1}, \boldsymbol{\Omega}_1^{-1})$, and Class 2 from $N(\boldsymbol{\mu}_{2}, \boldsymbol{\Omega}_{2}^{-1})$.
We set $\boldsymbol{\mu}_{1}=-\boldsymbol{\mu}_{2}=\left(0.5,0,0\right)$,
$\boldsymbol{\Omega}_{1}=\mathbf{I}_{3}-\boldsymbol{\Omega}$, and
$\boldsymbol{\Omega}_{2}=\mathbf{I}_{3}+\boldsymbol{\Omega}$, where
$\boldsymbol{\Omega}$ has entries $\omega_{22}=1$, $\omega_{11}=\omega_{33}=-0.60$, $\omega_{12}=\omega_{23}=-0.35$, and $\omega_{13}=0$. 
{\bf Left}: Scatterplot of $\left(X_{1},X_{2}\right)$ overlaid with corresponding theoretical contours for one simulated dataset.  
{\bf Right}: QDA classification error rate versus number of noise predictors. \label{fig:sim_toy}}
\end{figure}

To motivate later developments, we consider a two-class Gaussian classification problem with both linear and interaction effects with 3 true predictors. The oracle Bayes rule is to classify
an observation to class 1 if $Q\left(\mathbf{X}\right)>0$, and to
class 0 otherwise, where 
\begin{equation}
Q\left(\mathbf{X}\right)=1.627+X_{1}-0.6X_{1}^{2}-0.6X_{3}^{2}-0.7X_{1}X_{2}-0.7X_{2}X_{3}.\label{eq:toy}
\end{equation}
We simulated 100 independent datasets, each having $100$ observations in every class. Figure \ref{fig:sim_toy} shows the scatterplot of $\left(X_{1},X_{2}\right)$
for one simulated dataset. For each simulated dataset, we applied
LDA, logistic regression, and QDA to train classifiers, and the classification accuracy was estimated by using $1000$ additional testing samples generated from the Oracle model. As shown in Table \ref{tab:lda_vs_qda}, 
both LDA and logistic regression with only linear terms had poor prediction
powers, whereas QDA improved the classification accuracy dramatically. We further tested the classification accuracy of QDA when $k$ additional noise  predictors were included ($k=1, \ldots, 50$), each being drawn independently from $\mathcal{N}\left(0,1\right)$.
Figure \ref{fig:sim_toy} shows that the classification error rate of QDA
increased dramatically as the number of noise predictors increased, demonstrating the necessity of  developing methods capable of selecting both main effect and interaction terms efficiently.

\begin{table}[h]fa

\label{tab:lda_vs_qda}
\vspace{5pt}

\begin{centering}
\begin{tabular}{ccccc}
\hline 
\noalign{\vskip4bp}
Method & LDA & Logistic regression & QDA & QDA with 50 noise predictors\tabularnewline[\doublerulesep]
\noalign{\vskip\doublerulesep}
Test error \% & $34.81$ $(1.47)$ & $34.88$ $(1.38)$ & $15.65$ $(0.84)$ & $37.33$ $(1.78)$\tabularnewline[4bp]
\hline 
\noalign{\vskip\doublerulesep}
\end{tabular}
\par\end{centering}
\caption{Means (standard deviations) of testing error rates
for different classification methods over  $100$ replications. 
}\end{table}

 However, a direct application of Lasso on logistic
regression with all second-order terms is prohibitive for moderately
large $p$ (e.g., $p\ge1000$). To cope with this difficulty, \cite{fan2015innovated}
proposed innovated interaction screening (IIS) based on transforming
the original $p$-dimensional predictor vector by multiplying the
estimated precision matrix for each class. IIS first reduces the number
of predictors to a smaller order of $p$, and then identifies both
important main effects and interactions using the elastic net penalty
\citep{zou2005regularization}. The performance of the resulting method, IIS-SQDA,
relies heavily on the estimation of the $p\times p$ dimensional
precision matrix, which is usually a hard problem under high-dimensional
settings.  \cite{murphy2010variable}, \cite{zhang2011bic},
and \cite{maugis2011variable} proposed stepwise procedures for QDA
variable selection. These methods were shown to be consistent under
the multivariate Gaussian assumption on the design matrix. In
practice, however, performances of these methods can be much compromised when the normality assumption is violated, especially when predictors follow heavier-tailed distributions or when they are correlated in non-linear manners (see Section \ref{sec:simstudy}).

In order to gain robustness and computational efficiency,
we propose the method \underline{S}tepwise c\underline{O}nditional
likelihood variable selection for \underline{D}iscriminant \underline{A}nalysis (SODA) under the logistic regression framework, which starts
with a forward stepwise selection procedure to add in predictors with
main and/or interaction effects so as to reduce the number of candidate
predictors to a smaller order of $n$, and finishes with a backward
stepwise elimination procedure for further narrowing down individual
main and interaction effects. The criterion used for both forward addition and backward elimination is the extended Bayesian information criterion (EBIC) \citep{chen2008extended}. 

Although stepwise variable selection methods have been  widely known and used for regression problems, stepwise selection of interaction terms has been rare. Available methods typically consider adding interaction terms only among those predictors that have been selected for their main effects. In comparison, in each forward addition step, SODA evaluates the overall contribution of a predictor including both its main effects and its interactions with selected predictors. 
Under some regularity conditions, we
establish the screening consistency of the forward step  and 
the individual term selection consistency of the backward step of SODA
under high-dimensional settings. 

An interesting and powerful extension of SODA is for variable selection in multiple index models \citep{li1991sliced,cook2007fisher,jiang2014variable}, which assume that the response $Y$ (may be either discrete or continuous) depends on a subspace of $\mathbf{X}$ through an unknown (nonlinear) link function. The most popular method for estimating the subspace is the  sliced inverse regression (SIR) method \citep{li1991sliced}.  We note that after slicing (discretizing) the response variable $y$, we can apply SODA effectively for variable selection and model fitting. We call this extension the  \underline{S}liced \underline{SODA} (S-SODA).
Compared with variable selection methods based on SIR (see \cite{jiang2014variable} for references), S-SODA does not require the linearity condition  and enjoys much improved robustness without much sacrifice in sensitivity.

The rest of the article is organized as follows. SODA and S-SODA are presented in full detail in Section 2. Theoretical properties of SODA are studied in Section 3. Simulation results are shown in Section 4 to compare performances of SODA and S-SODA with those of other methods. In Section 5 We
further apply SODA to a couple of real examples to evaluate its empirical
performances, and in Section 6 conclude the article with a short discussion.
Detailed theoretical proofs and additional empirical results are provided in the Supplemental Materials.

\section{Variable and Interaction Selection for Discrete Response Models}

\subsection{Quadratic logistic regression model and its extended BIC}

We consider the $K$-class classification problem. Let $Y\in\left\{ 1,\dots,K\right\} $ denote the class label, let $\mathbf{X}=\left(X_{1},X_{2},\dots,X_{p}\right)^{T}$ be a vector of $p$ predictors, and let $\left\{ \left(\mathbf{x}_{i},y_{i}\right):\,i=1,\dots,n\right\}$
denote $n$ independent observations on $\left(\mathbf{X},Y\right)$.
When $p$ is large, usually only a small proportion of predictors
have predictive power on $Y$. Let $\mathcal{P}$ denote the set of
relevant predictors, and let $\mathcal{P}^{c}=\left\{ 1,\dots,p\right\} \backslash\mathcal{P}$ be noise ones. That is,
\[
P\left(Y\mid\mathbf{X}_{\mathcal{P}},\mathbf{X}_{\mathcal{P}^{c}}\right)=P\left(Y\mid\mathbf{X}_{\mathcal{P}}\right).
\]

We consider the following logistic model:
\begin{eqnarray}
p\left(Y=k\mid\mathbf{X},\boldsymbol{\theta}\right) & = & \frac{\exp\left[\delta_{k}\left(\mathbf{X}\mid\boldsymbol{\theta}\right)\right]}{1+\sum_{l=1}^{K-1}\exp\left[\delta_{l}\left(\mathbf{X}\mid\boldsymbol{\theta}\right)\right]},\quad k=1,\dots,K,\label{eq:main_model}
\end{eqnarray}
where $\delta_{k}\left(\mathbf{X}\mid\boldsymbol{\theta}\right)$
is the discriminant function for class $k$ and $\boldsymbol{\theta}$
denotes the vector of parameters. Choosing class $K$ as the baseline
class so that $\delta_{K}\left(\mathbf{X}\mid\boldsymbol{\theta}\right)=0$, we assume that
\begin{equation}
\delta_{k}\left(\mathbf{X}\mid\boldsymbol{\theta}\right)=\alpha_{k}+\boldsymbol{\beta}_{k}^{T}\mathbf{X}+\mathbf{X}^{T}\mathbf{A}_{k}\mathbf{X},\quad\mbox{for }k=1,\dots,K-1.
\label{eq:quadratic}
\end{equation}
Since $\mathbf{X}$ is conditioned on, we do not need to model the
distribution of $\mathbf{X}_{\mathcal{P}}$ or $\mathbf{X}_{\mathcal{P}^{c}}$,
which is both convenient and robust for variable selection.
Special cases of this model include:
\begin{itemize}
\item \vspace{1pt} Multinomial logistic regression (with $\mathbf{A}_{k}=\mathbf{0}$
for all $k$)
\item \vspace{-2pt}Linear/quadratic discriminant analysis, where $p\left(\mathbf{X}_{\mathcal{P}}\mid Y\right)$
is multivariate normal distribution
\item \vspace{-2pt}Discriminant analyses where $p\left(\mathbf{X}_{\mathcal{P}}\mid Y\right)$
is in the multivariate exponential family, 
\[
p\left(\mathbf{X}_{\mathcal{P}}=\mathbf{x}\mid Y=k,\boldsymbol{\eta}\right)=h\left(\mathbf{x}\right)g\left(\boldsymbol{\eta}_{k}\right)\exp\left(\boldsymbol{\eta}_{k}^{T}\mathbf{x}\right).
\]

\end{itemize}
To see the connection between QDA and model (\ref{eq:main_model}),
it is noted that for QDA models, 
\begin{eqnarray*}
\alpha_{k} & = & \log\left(\pi_{k}/\pi_{K}\right)-\frac{1}{2}\left(\log\left|\boldsymbol{\Sigma}_{k}\right|-\log\left|\boldsymbol{\Sigma}_{K}\right|+\boldsymbol{\mu}_{k}^{T}\boldsymbol{\Sigma}_{k}^{-1}\boldsymbol{\mu}_{k}-\boldsymbol{\mu}_{K}^{T}\boldsymbol{\Sigma}_{K}^{-1}\boldsymbol{\mu}_{K}\right),\\
\boldsymbol{\beta}_{k}^{T} & = & \boldsymbol{\mu}_{k}^{T}\boldsymbol{\Sigma}_{k}^{-1}-\boldsymbol{\mu}_{K}^{T}\boldsymbol{\Sigma}_{K}^{-1},\\
\mathbf{A}_{k} & = & -\frac{1}{2}\left(\boldsymbol{\Sigma}_{k}^{-1}-\boldsymbol{\Sigma}_{K}^{-1}\right),\quad\mbox{for }k=1,\dots,K-1.
\end{eqnarray*}

Let $\mathcal{M}$ and $\mathcal{I}$ denote subsets of main effects
and interaction pairs, respectively, and let $\mathcal{M}_{0}$ and
$\mathcal{I}_{0}$ denote the corresponding true sets defined as  
\[
\mathcal{M}_{0}=\left\{ j:\,\exists\,k\,\,\mbox{ s.t. }\,\,\beta_{k,j}\neq0\right\} \quad\mbox{and}\quad\mathcal{I}_{0}=\left\{ \left(i,j\right):\,\exists k\,\,\mbox{ s.t. }\,A_{k,i,j}\neq0\right\} ,
\]
with $k$ indicating the class label. Let $\mathcal{A}=\mathcal{M}_{0}\cup\mathcal{I}_{0}$ denote the true  set of all effects, and let $\mathcal{S}=\mathcal{M}\cup\mathcal{I}$. The true set of relevant predictors $\mathcal{P}$ can be derived
from $\mathcal{A}$ as 
\[
\mathcal{P}=\mathcal{M}_{0}\cup\left\{ j:\,\exists\,i\:\:\mbox{s.t.}\,\,\left(i,j\right)\in\mathcal{I}_{0}\right\} .
\]
Our main objective is to infer $\mathcal{A}$, with a special interest in terms in  $\mathcal{I}_{0}$.

Let $\boldsymbol{\theta}_{\mathcal{S}}$ denote the collection of all coefficients in model (\ref{eq:quadratic}),
whose $0$'s correspond to terms not in $\mathcal{S}$, and let $\boldsymbol{\theta}_{k,\mathcal{S}}$
denote the corresponding coefficients for class $k$. For a dataset
$\left\{ \left(\mathbf{x}_{i},y_{i}\right):\,i=1,\dots,n\right\} $,
the log-likelihood for $\boldsymbol{\theta}_{\mathcal{S}}$ is denoted
as $l_{n}\left(\boldsymbol{\theta}_{\mathcal{S}}\right)$. Let $\mathbf{Z} \equiv (1,\mathbf{X}, \mathbf{X}\bigotimes \mathbf{X})$
be the augmented version of $\mathbf{X}$, containing intercept $1$,
main effects, and all interaction terms of $\mathbf{X}$. Let $\mathbf{z}_{i}$
be the $i$-th observation of $\mathbf{Z}$. Then $l_{n}\left(\boldsymbol{\theta}_{\mathcal{S}}\right)$
takes the form of a logistic regression model in $\mathbf{Z}$: 
\[
l_{n}\left(\boldsymbol{\theta}_{\mathcal{S}}\right)\:=\:\sum_{i=1}^{n}\left\{ \boldsymbol{\theta}_{y_{i},\mathcal{S}}^{T}\mathbf{z}_{i}-\log\left(1+\sum_{l=1}^{K-1}\exp\left(\boldsymbol{\theta}_{l,\mathcal{S}}^{T}\mathbf{z}_{i}\right)\right)\right\} .
\]

Let $\tilde{\boldsymbol{\theta}}_{\mathcal{S}}$ denote the MLE of
$\boldsymbol{\theta}_{\mathcal{S}}$. By Lemma 2 in the appendix,
with high probability $l_{n}\left(\boldsymbol{\theta}_{\mathcal{S}}\right)$
is convex and $\tilde{\boldsymbol{\theta}}_{\mathcal{S}}$ can be
obtained by Newton-Raphson algorithm. Let $\boldsymbol{\theta}_{0}$
denote the true parameter vector. Theorem 1 illustrates the consistency
of $\tilde{\boldsymbol{\theta}}_{\mathcal{S}}$ for any reasonable set $\mathcal{S}$.
\begin{thm}
\textbf{\emph{\label{theorem:MCLE_asymptotics}}}Under Conditions
C1 $\sim$ C4 in Section \ref{sec:theory}, as $n\to\infty$,
\begin{equation}
\max_{\mathcal{S}\supset\mathcal{A},\,\left|\mathcal{S}\right|\le Q}\left\Vert \tilde{\boldsymbol{\theta}}_{\mathcal{S}}-\boldsymbol{\theta}_{0}\right\Vert _{2}=O_{p}\left(n^{-1/2+\xi}\right),
\end{equation}
for any constants $0<\xi<1/2$ and $Q\ge\left|\mathcal{A}\right|$ independent of  $n$.
\end{thm}
In high-dimensional settings, the classic Bayesian information criterion (BIC) \citep{schwarz1978estimating} is too liberal and tends
to select many false positives \citep{broman2002model}. \cite{chen2008extended}
proposed extended BIC (EBIC) and showed it to be consistent for linear
regression models under high-dimensional settings. The EBIC for set
$\mathcal{S}$ is specified as 
\begin{equation}
\text{EBIC}_{\gamma}\left(\mathcal{S}\right)\,=\,-2\,l_{n}\left(\tilde{\boldsymbol{\theta}}_{\mathcal{S}}\right)+\left|\mathcal{S}\right|\log n+2\gamma\,\left|\mathcal{S}\right|\log p,\label{eq:EBIC}
\end{equation}
where $\left|\mathcal{S}\right|$ is the size of set $\mathcal{S}$,
and $\gamma$ is a tuning parameter. 
The selection of $\gamma$ may depend on the relative sizes of $n$ and $p$, and 
some heuristics on determining $\gamma$ practically is discussed in section \ref{sub:impl}. Let $\tilde{\mathcal{S}}_{\text{EBIC}}$
be the selected set of predictors minimizing the EBIC, and let $Q$ be
any positive constant greater than constant $p_{0}$ in condition
(C1) in section \ref{sec:theory}. Then,
\begin{equation}
\tilde{\mathcal{S}}_{\text{EBIC}}\,=\,\underset{\mathcal{S}:\:\left|\mathcal{S}\right|\,\leq\,Q}{\arg\min}\;\text{EBIC}_{\gamma}\left(\mathcal{S}\right),\label{eq:S_hat}
\end{equation}
where $\left|\mathcal{S}\right|$ denotes the size of set $\mathcal{S}$.
The asymptotic property of $\tilde{\mathcal{S}}_{\text{EBIC}}$ is
shown by the following theorem.
\begin{thm}
\textbf{\emph{(EBIC criterion consistency)}}\label{theorem:EBIC}
Under Conditions C1 $\sim$ C4 in Section \ref{sec:theory}, $\tilde{\mathcal{S}}_{\text{EBIC}}$
is a consistent estimator of $\mathcal{A}$, i.e.,
\[
\text{Pr}\left(\tilde{\mathcal{S}}_{\text{EBIC}}=\mathcal{A}\right)\to1,\mbox{ as }n\to\infty,
\]
for any $\gamma>2-1/\left(2\kappa\right)$.
\end{thm}
By treating our model as a logistic regression on $\left(\mathbf{Z},Y\right)$,
Theorem \ref{theorem:EBIC} follows directly from the asymptotic consistency
of EBIC for generalized linear models (GLM), which was proved in \cite{chen2012extended}
and \cite{foygel2011bayesian} in both fixed and random design contexts.
We thus omit its proof. Different from \cite{chen2012extended} and
\cite{foygel2011bayesian}, here we require $\gamma>2-1/\left(2\kappa\right)$
instead of $\gamma>1-1/\left(2\kappa\right)$ to penalize additional
model flexibility caused by the inclusion of interaction terms.

\subsection{SODA: a stepwise variable and interaction selection procedure}

In practice it is infeasible to enumerate all possible $\mathcal{S}$
to find the one that minimizes the EBIC. For a closely related generalized
linear model variable selection problem, \cite{chen2012extended}
and \cite{foygel2011bayesian} used Lasso \citep{tibshirani1996regression}
to obtain a solution path of predictor sets, and chose the optimal
set on the path with the lowest EBIC. However, this method also fails under the
high-dimensional setting for QDA, in which there are $O\left(p^{2}\right)$
candidate interaction terms. Furthermore, Lasso's variable
selection consistency for logistic regression requires the incoherence
condition  \citep{ravikumar2010high}, which can be easily violated due
to correlations between interaction terms and their corresponding
main effect terms. The IIS procedure proposed in \cite{fan2015innovated}
requires the estimation of the $p\times p$ precision matrix, which is
by itself a challenging problem. If the related and unrelated predictors
are moderately correlated, IIS\textquoteright s marginal screening
strategy has difficulties in filtering out noise predictors.
We propose here the stepwise
procedure SODA, consisting of three stages: (1) a preliminary forward main effect
selection; (2) forward variable selection (considering both main and interaction effects), and (3) backward elimination. 
\begin{enumerate}
\item \textbf{Preliminary main effect selection}: This step is the same as that in the standard stepwise regression method. Let $\mathcal{M}_{t}$ denote
the selected set of main effects at step $t$. SODA starts with $\mathcal{M}_{1}=\emptyset$
and iterates the operations below until termination.

\begin{enumerate}
\item \vspace{-5pt}For each predictor $j\notin\mathcal{M}_{t}$, create
a new candidate set $\mathcal{M}_{t,j}=\mathcal{M}_{t}\cup\left\{ j\right\} $.
\item \vspace{2pt}Find the predictor $j$ with lowest $\mbox{EBIC}_{\gamma}\left(\mathcal{M}_{t,j}\right)$.
If $\mbox{EBIC}_{\gamma}\left(\mathcal{M}_{t,j}\right)<\mbox{EBIC}_{\gamma}\left(\mathcal{M}_{t}\right)$,
continue with $\mathcal{M}_{t+1}=\mathcal{M}_{t,j}$, otherwise terminate
with $\tilde{\mathcal{M}}_{F}$ and go to 2.
\end{enumerate}
\item \textbf{Forward variable addition (both main and interaction effects):} Let $\mathcal{C}_{t}$ denote
the selected set of predictors at step $t$, and let $\mathcal{S}_{t}=\tilde{\mathcal{M}}_{F}\cup\mathcal{C}_{t}\cup\left(\mathcal{C}_{t}\times\mathcal{C}_{t}\right)$
denote the set of terms induced by $\mathcal{C}_{t}$. SODA starts
with $\mathcal{C}_{1}=\emptyset$ and iterate the operations below
until termination.

\begin{enumerate}
\item \vspace{-5pt}For each $j\notin\mathcal{C}_{t}$, create a candidate
set $\mathcal{C}_{t,j}=\mathcal{C}_{t}\cup\left\{ j\right\} $ and
let $\mathcal{S}_{t,j}=\tilde{\mathcal{M}}_{F}\cup\mathcal{C}_{t,j}\cup\left(\mathcal{C}_{t,j}\times\mathcal{C}_{t,j}\right)$.
\item \vspace{2pt}Find the predictor $j$ with lowest $\mbox{EBIC}_{\gamma}\left(\mathcal{S}_{t,j}\right)$.
If $\mbox{EBIC}_{\gamma}\left(\mathcal{S}_{t,j}\right)<\mbox{EBIC}_{\gamma}\left(\mathcal{S}_{t}\right)$,
continue with $\mathcal{C}_{t+1}=\mathcal{C}_{t,j}$, otherwise terminate
with $\tilde{\mathcal{C}}_{F}$ and go to 3.
\end{enumerate}
\item \textbf{Backward elimination}: Let $\mathcal{S}_{t}$ denote the selected
set of individual terms at step $t$ of backward stage. SODA starts
with $\mathcal{S}_{1}=\tilde{\mathcal{M}}_{F}\cup\tilde{\mathcal{C}}_{F}\cup\left(\tilde{\mathcal{C}}_{F}\times\tilde{\mathcal{C}}_{F}\right)$
and iterate the operations below until termination.

\begin{enumerate}
\item \vspace{-5pt}For each main or interaction term $j\in\mathcal{S}_{t}$
(e.g. $j=1$ or $j=\left(1,2\right)$), create a candidate set $\mathcal{S}_{t,j}=\mathcal{S}_{t}\backslash\left\{ j\right\} $.
\item \vspace{2pt}Find term $j$ with lowest $\mbox{EBIC}_{\gamma}\left(\mathcal{S}_{t,j}\right)$.
If $\mbox{EBIC}_{\gamma}\left(\mathcal{S}_{t,j}\right)<\mbox{EBIC}_{\gamma}\left(\mathcal{S}_{t}\right)$,
remove term $j$, otherwise terminate and retain set $\tilde{\mathcal{S}}=\mathcal{S}_{t}$.
\end{enumerate}
\end{enumerate}

Stepwise methods had been primary tools for conducting variable selection in regression problems long before the recent development of Lasso-type methods. The forward stepwise procedure has also been considered for
 variable screening for linear regressions  in high-dimensional settings \citep{wasserman2009high,wang2009forward}. When considering interactions, a standard approach typically examines only those among the variables that have been deemed significant due to their main effects.  
However, Stage 2 of SODA for forward variable addition is different.
After the preliminary selection of Stage 1, in Stage 2 SODA keeps track of a new set of variables ${\cal C}_t$, of which all main and quadratic terms are considered together. In other words, at each step SODA  evaluates the EBIC for the overall effect of adding one predictor. Thus, choosing one variable to add in the forward variable selection stage is of order $O(p)$, and the whole stage is of order $O(ps)$, where $s$ is the number of truly relevant predictors. A naive method that searches through all individual terms is of order $O\left(p^{2} s^2 \right)$. 
Another important feature of SODA is that each  backward step only eliminates one individual term instead of all terms related to one predictor.
In other words, SODA selects individual main and interaction effect
terms without any nesting requirements. 

Our theory shows that the forward variable addition step is sufficient 
for SODA to achieve the screening consistency. However, the number of parameters
and the EBIC penalization in this forward step
increases quadratically with the cardinality of $\mathcal{C}_{t}$.
Therefore it can be hard to add predictors with only weak main effects.
To optimize the empirical performance, we include the preliminary
main effect selection stage to identify predictors with only weak
main effects. 

\subsection{Sliced SODA (S-SODA) for general index models}
In his seminal work on nonlinear dimension reduction, \cite{li1991sliced} proposed
a semi-parametric index model of the form
\begin{equation}
Y=f\left(\boldsymbol{\beta}_{1}^{T}\mathbf{X},\boldsymbol{\beta}_{2}^{T}\mathbf{X},\dots,\boldsymbol{\beta}_{d}^{T}\mathbf{X},\varepsilon\right),\label{eq:gim}
\end{equation}
where $f$ is an unknown function and $\varepsilon$ is random error
independent of $\mathbf{X}$, and  the sliced
inverse regression (SIR) method to estimate the central dimension
reduction subspace (CDRS) spanned by the directions $\boldsymbol{\beta}_{1},\dots,\boldsymbol{\beta}_{d}$.
Since the estimation of CDRS does not automatically lead
to variable selection, several methods have been developed to do simultaneous 
dimension reduction and variable selection  for index
models. For example, \cite{li2005model} designed a backward subset
selection method, and \cite{li2007sparse}
developed the sparse SIR (SSIR) algorithm to obtain shrinkage estimates
of the SDR directions under $L_{1}$ norm.  Motivated by stepwise regression for linear models, \cite{zhong2012correlation}
proposed a forward stepwise variable selection procedure called correlation
pursuit (COP) for index models. \cite{lin2015} showed the necessary and sufficient condition for SIR to be consistent in high-dimensional settings and introduced a diagonal thresholding method, DT-SIR, for variable selection. \cite{lin2016} proposed a new formulation of the SIR estimation and a direct  application of Lasso for variable selection with index models.

The aforementioned SIR-based methods  consider primarily the information from the first conditional moment, $\mathbb{E}\left(\mathbf{X}\mid Y\right)$,
and tend to miss important variables with second-order
effects. In order to overcome this problem, \cite{jiang2014variable}
proposed SIRI, which utilizes both the first and the second conditional
moments  to select variables. SIRI derives its   procedure
from a likelihood-ratio test perspective by assuming the following 
working model:  
\begin{equation}
\begin{aligned}\mathbf{X}_{\mathcal{P}}\mid s\left(Y\right)=h & \;\:\sim\;\:\mathcal{N}\left(\boldsymbol{\mu}_{h},\boldsymbol{\Sigma}_{h}\right),\\
\mathbf{X}_{\mathcal{P}^{c}}\mid\mathbf{X}_{\mathcal{P}},s\left(Y\right)=h & \;\:\sim\;\:\mathcal{N}\left(\boldsymbol{a}+\boldsymbol{B}^{T}\mathbf{X}_{\mathcal{P}},\boldsymbol{\Sigma}_{0}\right),
\end{aligned}
\quad h=1,\dots,H.\label{eq:SIRI}
\end{equation}
where  $\mathcal{P}$ denote the set of true predictors.
\cite{jiang2014variable} showed that SIRI is a consistent
variable selection procedure for model (\ref{eq:SIRI}), and also for a more general class of models satisfying the following linearity and constant variance conditions. In fact, all the aforementioned methods  
require either the linearity condition or the constant variance condition, or both.

\vspace{3mm}

\noindent
\textbf{Linearity condition:} $E\left(\mathbf{X}_{\mathcal{P}^{c}}\mid\mathbf{X}_{\mathcal{P}}\right)$
is linear in $\mathbf{X}_{\mathcal{P}}$.

\noindent
\textbf{Constant variance condition:} $\text{Cov}\left(\mathbf{X}_{\mathcal{P}^{c}}\mid\mathbf{X}_{\mathcal{P}}\right)$
is a constant.

\vspace{3mm}

When the linearity  and constant variance conditions approximately
hold, SIRI and other SIR-related methods usually enjoy excellent empirical performances.
However, when either  condition is violated, the performances of these methods deteriorate rapidly.  This issue motivates us to develop  sliced-SODA (S-SODA), a modification of  SODA. As a working model, S-SODA can be seen as assuming only 
the first half of model (\ref{eq:SIRI}) without any distributional
assumption on $\mathbf{X}_{\mathcal{P}^{c}}$:
\begin{equation}
\begin{aligned}\mathbf{X}_{\mathcal{P}}\mid s\left(Y\right)=h & \;\:\sim\;\:\mathcal{N}\left(\boldsymbol{\mu}_{h},\boldsymbol{\Sigma}_{h}\right), 
\end{aligned}
\quad h=1,\dots,H.\label{eq:S-SODA_model}
\end{equation}
Note that model (\ref{eq:S-SODA_model}) is essentially the QDA model,
and we can apply SODA as a consistent variable selection procedure.
More precisely,  S-SODA starts by
sorting the samples in ascending order of $y_{i}$, and  equally
partitioning them into $H$ slices (the discretization step). It then applies SODA to  data $\left\{ \left(s_{i},\mathbf{x}_{i}\right)\right\} _{i=1}^{n}$, where   $s_{i}$ denote the slice index for $y_{i}$. S-SODA finally outputs all the selected main and interaction terms.

\subsection{Post-selection prediction for continuous response}

S-SODA conducts variable selection for semi-parametric model (\ref{eq:gim})  without knowing the true functional form of the link function. After variable selection, it is of interest to predict the response
variable $\tilde{y}$ for a new observation of predictors $\tilde{\mathbf{x}}$.
Suppose our training data consist of $n$ independent observations $\left\{ \left(y_{i},\mathbf{x}_{i}\right)\right\} _{i=1}^{n}$. 
Let $\tilde{\mathcal{S}}$ denote the selected set of terms by S-SODA, and let
$\tilde{\mathcal{P}}$ denote the set of predictors with any term
in $\tilde{\mathcal{S}}$, which is the S-SODA estimate of $\mathcal{P}$.
Let $\hat{\boldsymbol{\mu}}=\left(\hat{\boldsymbol{\mu}}_{1},\dots,\hat{\boldsymbol{\mu}}_{H}\right)$,
$\hat{\boldsymbol{\Sigma}}=\left(\hat{\boldsymbol{\Sigma}}_{1},\dots,\hat{\boldsymbol{\Sigma}}_{H}\right)$,
where $\hat{\boldsymbol{\mu}}_{h}$ and $\hat{\boldsymbol{\Sigma}}_{h}$
are respectively the sample mean vector and covariance matrix of $\mathbf{X}_{\tilde{\mathcal{P}}}$
in slice $h$. Note that $\hat{\boldsymbol{\mu}}$ and $\hat{\boldsymbol{\Sigma}}$
are MLEs of parameters in model (\ref{eq:S-SODA_model}). Inverting
model (\ref{eq:S-SODA_model}) by the Bayes rule, we have 
\[
\text{Pr}\left(s\left(Y\right)=h\mid\mathbf{X}_{\tilde{\mathcal{P}}},\boldsymbol{\mu},\boldsymbol{\Sigma}\right)=\frac{N\left(\mathbf{X}_{\tilde{\mathcal{P}}}\mid\boldsymbol{\mu}_{h},\boldsymbol{\Sigma}_{h}\right)}{\sum_{l=1}^{H}N\left(\mathbf{X}_{\tilde{\mathcal{P}}}\mid\boldsymbol{\mu}_{l},\boldsymbol{\Sigma}_{l}\right)},\quad h=1,\dots,H.
\]

We consider the conditional expectation $\mathbb{E}\left[Y\mid\mathbf{X}_{\tilde{\mathcal{P}}}\right]$
as prediction of $Y$ given $\mathbf{X}_{\tilde{\mathcal{P}}}$. Note
that 
\begin{eqnarray*}
\mathbb{E}\left[Y\mid\mathbf{X}_{\tilde{\mathcal{P}}}\right] & = & \sum_{h=1}^{H}\mathbb{E}\left[Y\mid s\left(Y\right)=h,\mathbf{X}_{\tilde{\mathcal{P}}}\right]\text{Pr}\left(s\left(Y\right)=h\mid\mathbf{X}_{\tilde{\mathcal{P}}},\boldsymbol{\mu},\boldsymbol{\Sigma}\right)\\
 & = & \sum_{h=1}^{H}\frac{\mathbb{E}\left[Y\mid s\left(Y\right)=h,\mathbf{X}_{\tilde{\mathcal{P}}}\right]\cdot N\left(\mathbf{X}_{\tilde{\mathcal{P}}}\mid\boldsymbol{\mu}_{h},\boldsymbol{\Sigma}_{h}\right)}{\sum_{l=1}^{H}N\left(\mathbf{X}_{\tilde{\mathcal{P}}}\mid\boldsymbol{\mu}_{l},\boldsymbol{\Sigma}_{l}\right)}.
\end{eqnarray*}
We use a plug-in estimator of $\mathbb{E}\left[Y\mid\mathbf{X}_{\tilde{\mathcal{P}}}\right]$,
denoted as $\hat{Y}=\hat{\mathbb{E}}\left[Y\mid\mathbf{X}_{\tilde{\mathcal{P}}}\right]$,
where
\begin{equation}
\hat{Y}=\hat{\mathbb{E}}\left[Y\mid\mathbf{X}_{\tilde{\mathcal{P}}}\right]=\sum_{h=1}^{H}\frac{\hat{M}_{h}\cdot N\left(\mathbf{X}_{\tilde{\mathcal{P}}}\mid\hat{\boldsymbol{\mu}}_{h},\hat{\Sigma}_{h}\right)}{\sum_{l=1}^{H}N\left(\mathbf{X}_{\tilde{\mathcal{P}}}\mid\hat{\boldsymbol{\mu}}_{l},\hat{\Sigma}_{l}\right)}.\label{eq:Y_pred}
\end{equation}
where $\hat{M}_{h}$ is the sample mean of response $Y$ in slice
$h$. $\hat{M}_{h}$ can be considered as the zero-th order approximation
to $\mathbb{E}\left[Y\mid s\left(Y\right)=h,\mathbf{X}_{\tilde{\mathcal{P}}}\right]$,
in the sense that $\hat{M}_{h}$ is independent of $\mathbf{X}_{\tilde{\mathcal{P}}}$.
A more sophisticated model is to consider the first-order approximation
that models $\mathbb{E}\left[Y\mid s\left(Y\right)=h,\mathbf{X}_{\tilde{\mathcal{P}}}\right]$
as a linear combination of $\mathbf{X}_{\tilde{\mathcal{P}}}$ in
each slice.

\subsection{Implementation issues with SODA: tuning parameter and screening depth\label{sub:impl}}

Sections \ref{sec:theory} characterizes asymptotic
properties of the EBIC and SODA and provides some guidelines for
choosing the tuning parameter
$\gamma$ of EBIC. However, these asymptotic results are not directly
usable . In practice, we propose to use a $10$-fold cross-validation
(CV) procedure for selecting $\gamma$ from $\left\{ 0,0.5,1.0\right\}$.
For simulation studies and real data analyses in Sections \ref{sec:simstudy}
and \ref{sec:realdata}, to make SODA more easily comparable with
Lasso-EBIC studied in \cite{chen2012extended}, we fixed $\gamma=0.5$
as suggested in \cite{chen2012extended}.

The forward variable addition stage terminates if EBICs of all candidate
models are larger than the EBIC of the current model. Therefore, the
screening depth of the forward stage is determined by the EBIC. In
Theorem \ref{theorem:forward}, we show that this procedure is asymptotically
screening consistent; namely, the truly relevant terms will be all included
by the end of the forward stage. Nevertheless, SODA is not sensitive
to adding more terms in the forward stage since those unrelated terms
will be eventually eliminated in the backward stage. Missing
one relevant term is usually more harmful than including one noise
term. Therefore, to optimize the empirical performance, we let SODA
continue the forward variable addition  for  $p_{f}$ steps
after the step that fails to decrease EBIC (default $p_{f}=3$).

\section{Theoretical Properties of SODA\label{sec:theory}}

To study theoretical properties of SODA procedure, we assume the following
conditions:
\begin{enumerate}
\item[\textbf{(C1)}] The divergence speed of $p$ is bounded above by $p\le n^{\kappa}$
for some $\kappa>0$, and the size of the true predictor set $\mathcal{P}$
is bounded as $\left|\mathcal{P}\right|\le p_{0}$ for a fixed integer
$p_{0}$.

\item[\textbf{(C2)}] Magnitudes of true coefficients in $\boldsymbol{\theta}_{\mathcal{A}}$
are bounded above and below by constants, namely there exist positive
constants $\theta_{\max}>\theta_{\min}>0$ such that 
\[
\theta_{\min}\le\min\left\{ \left|\theta_{j}\right|:\,j\in\mathcal{A}\right\} \le\max\left\{ \left|\theta_{j}\right|:\,j\in\mathcal{A}\right\} \le\theta_{\max}.
\]

\item[\textbf{(C3)}] Let $\mathbf{Z}$ be the augmented version of $\mathbf{X}$,
containing intercept $1$, as well as all first-order and second-order
terms of $\mathbf{X}$. Each $Z_{j}\in\mathbf{Z}$ is sub-exponential,
i.e. there are positive constants $C_{1}$ and $C_{2}$ such that,
\[
\text{Pr}\left(\left|Z_{j}-\mathbb{E}\left[Z_{j}\right]\right|>t\right)\le C_{1}\exp\left(-C_{2}t\right)\quad\text{ for all }t>0.
\]

\item[\textbf{(C4)}] Let $\text{Cov}\left(\mathbf{Z}\right)$ denote the
covariance matrix of $\mathbf{Z}$. There exist constants $0<\tau_{1}<\tau_{2}<\infty$
such that 
\[
\tau_{1}\le\lambda_{\min}\left(\text{Cov}\left(\mathbf{Z}\right)\right)<\lambda_{\max}\left(\text{Cov}\left(\mathbf{Z}\right)\right)\le\tau_{2},
\]
where $\lambda_{\min}\left(\cdot\right)$ and $\lambda_{\max}\left(\cdot\right)$
denote the smallest and largest eigenvalues of a matrix.

\end{enumerate}


We show that the forward variable addition stage (Stage 2 of SODA) is
already screening consistent. To proceed, we need to define the following
concept to study the stepwise detectability of true predictors in
$\mathcal{P}$. 
Let $\boldsymbol{\theta}_{\mathcal{S}}^{*}$ denote the population
version of the risk minimizer, 
\[
\boldsymbol{\theta}_{\mathcal{S}}^{*}=\underset{\boldsymbol{\theta}_{\mathcal{S}}}{\arg\min}\,\mathbb{E}\left[-\log p\left(Y\mid\mathbf{X},\boldsymbol{\theta}_{\mathcal{S}}\right)\right],
\]
where the expectation is over the joint distribution of $\left(Y,\mathbf{X}\right)$.
Let vector $\boldsymbol{\theta}_{\mathcal{S}}^{j*}$ be parameters
in $\boldsymbol{\theta}_{\mathcal{S}}^{*}$ associated with predictor
$X_{j}$. The stepwise detectable condition is necessary for the screening consistency of the forward variable addition stage.
\begin{defn}
\textbf{(Stepwise detectable condition)} A set of predictors $\mathcal{C}_{1}$
is stepwise detectable given $\mathcal{C}_{2}$ if $\mathcal{C}_{1}\cap\mathcal{C}_{2}=\emptyset$,
and for any set $\mathcal{C}$ satisfying $\mathcal{C}\supset\mathcal{C}_{2}$
and $\mathcal{C}\not\supset\mathcal{C}_{1}$, there exist constants
$\theta_{\max}>\theta_{\min}>0$, such that
\[
\theta_{\min}\le\max_{j\in\mathcal{C}^{c}\cap\mathcal{C}_{1}}\left\Vert \boldsymbol{\theta}_{\mathcal{S}_{\mathcal{C}\cup\left\{ j\right\} }}^{j*}\right\Vert _{\infty}\le\theta_{\max},
\]
where $\mathcal{S}_{\mathcal{C}\cup\left\{ j\right\} }=\mathcal{M}^{j}\cup\mathcal{I}^{j}$
with $\mathcal{M}^{j}=\mathcal{C}\cup\left\{ j\right\} $ and $\mathcal{I}^{j}=\mathcal{M}^{j}\times\mathcal{M}^{j}$,
and $\left\Vert \cdot\right\Vert _{\infty}$ denotes the $L_{\infty}$
norm. Let $\mathcal{T}_{m}=\left\{ j:\,\text{predictor }j\mbox{ is stepwise detectable given }\cup_{i=0}^{m-1}\mathcal{T}_{i-1}\right\} $
and $\mathcal{T}_{0}=\emptyset$. The set of true predictors $\mathcal{P}$
is said to be stepwise detectable if $j\in\cup_{i=1}^{\infty}\mathcal{T}_{i}$
for all $j\in\mathcal{P}$.
\end{defn}
In other words, if the current selection $\mathcal{C}$ contains $\mathcal{C}_{2}$,
then there always exist detectable predictors conditioning on currently
selected variables until we include all the predictors indexed by
$\mathcal{C}_{1}$. A true predictor $j\in\mathcal{P}$ is not stepwise
detectable either because it perfectly correlates with some other
terms, or its effects can only be detected conditioning on some other
stepwise undetectable terms.

We give an example to illustrate the scenarios when stepwise detectable
condition may or may not hold. Suppose there are two true jointly
normal relevant predictors $X_{1}$ and $X_{2}$ with means $\mu_{1}$
and $\mu_{2}$, and there is only one interaction term $X_{1}X_{2}$
in model (\ref{eq:main_model}), i.e. $\mathcal{A}=\left\{ \left(1,2\right)\right\} $.
$\mathcal{P}$ is not stepwise detectable if both $\mu_{1}=0$ and
$\mu_{2}=0$. Starting from empty set $\emptyset$, the forward procedure
will not add $X_{1}$ or $X_{2}$ into the model, because there is
no main effect for $X_{1}$ and $X_{2}$ and the interaction term
$X_{1}X_{2}$ does not correlate with marginal terms $X_{1}$ and
$X_{2}$ ($\text{Cov}\left(X_{1},X_{1}X_{2}\right)=0$ and $\text{Cov}\left(X_{2},X_{1}X_{2}\right)=0$).
However, if either $\mu_{1}\neq0$ or $\mu_{2}\neq0$, $\mathcal{P}=\left\{ 1,2\right\} $
is stepwise detectable.

Let $\tilde{\mathcal{S}}_{F}=\tilde{\mathcal{M}}_{F}\cup\tilde{\mathcal{C}}_{F}\cup\left(\tilde{\mathcal{C}}_{F}\times\tilde{\mathcal{C}}_{F}\right)$
denote the selected set of terms at the end of forward variable addition stage.
It is unrealistic to require $\tilde{\mathcal{S}}_{F}=\mathcal{A}$.
However, it should be demanded that $\tilde{\mathcal{S}}_{F}\supseteq\mathcal{A}$,
i.e. $\tilde{S}_{F}$ contains all relevant terms. We define the forward
stage to be screening consistent if $p\left(\tilde{\mathcal{S}}_{F}\supseteq\mathcal{A}\right)\to1$.
We also do not want the size of $\tilde{\mathcal{S}}_{F}$ to be too
large, otherwise forward variable addition loses its purpose. The screening
consistency of forward stage is established by the following theorem.
\begin{thm}
\emph{\label{theorem:forward}}\textbf{\emph{(Forward stage screening
consistency)}}\emph{ }If conditions C1 $\sim$ C4 hold, and all predictors
in $\mathcal{P}$ are stepwise detectable, then the forward variable addition stage finishes in finite number of steps and is screening
consistent. In particular, as $n\to\infty$,
\[
\text{Pr}\left(\left|\tilde{\mathcal{C}}_{F}\right|\le Q\right)\to1,\;\,\mbox{ and }\;\,\text{Pr}\left(\tilde{\mathcal{C}}_{F}\supseteq\mathcal{P}\right)\to1,
\]
where $Q=\left\lceil 8\lambda_{1}^{-1}\theta_{\min}^{-2}\log K\right\rceil $,
$\lambda_{1}$ is a positive constant defined in Lemma 2 in appendix,
$K$ is the number of classes, and $\theta_{\min}$ is a positive
constant defined in condition C2.
\end{thm}
In other words, asymptotically $\tilde{\mathcal{C}}_{F}$ contains
all predictors in $\mathcal{P}$, which implies $\tilde{\mathcal{S}}_{F}\supseteq\mathcal{A}$,
and the forward stage stops in finite number of steps.
We show in the following theorem two uniform bounds guaranteeing that
all unrelated terms will be eliminated and all related terms will
be kept in the backward stage.
\begin{thm}
\label{theorem:backward}\textbf{\emph{(Uniform bound of EBIC in backward
stage)}}\emph{ }Fix any positive constant $Q>0$. Under conditions
C1 $\sim$ C4, as $n\to\infty$,\emph{ }
\begin{equation}
\text{Pr}\left(\max_{\mathcal{S}\supsetneq\mathcal{A}:\left|\mathcal{S}\right|\le Q}\min_{j\in\mathcal{S}\backslash\mathcal{A}}\left\{ \text{EBIC}_{\gamma}\left(\mathcal{S}\backslash\left\{ j\right\} \right)-\text{EBIC}_{\gamma}\left(\mathcal{S}\right)\right\} <0\right)\to1,\label{eq:backward_1}
\end{equation}
and
\begin{equation}
\text{Pr}\left(\min_{\mathcal{S}\supset\mathcal{A}:\left|\mathcal{S}\right|\le Q}\min_{j\in\mathcal{A}}\left\{ \text{EBIC}_{\gamma}\left(\mathcal{S}\backslash\left\{ j\right\} \right)-\text{EBIC}_{\gamma}\left(\mathcal{S}\right)\right\} <0\right)\to0,\label{eq:backward_2}
\end{equation}
for any constant $\gamma>Q-\left|\mathcal{A}\right|-\left(2\kappa\right)^{-1}$.
\end{thm}
Eq (\ref{eq:backward_1}) implies that if $\mathcal{S}\supsetneq\mathcal{A}$
and $\left|\mathcal{S}\right|\le Q$, there will be at least one unrelated
term $j\in\mathcal{S}\cap\mathcal{A}^{c}$ such that removing $j$
from $\mathcal{S}$ leads to lower EBIC. Eq (\ref{eq:backward_2})
implies that if $\mathcal{S}\supset\mathcal{A}$ and $\left|\mathcal{S}\right|\le Q$,
there is no related term $j\in\mathcal{A}$ such that removing $j$
from $\mathcal{S}$ leads to lower EBIC. As a summary, as $n\to\infty$,
with probability tending to $1$, no related term will be eliminated,
and all unrelated terms will be eliminated in the backward stage until
$\tilde{\mathcal{S}}=\mathcal{A}$. Theorem \ref{theorem:backward}
requires candidate sets have finite size ($\left|\mathcal{S}\right|\le Q$),
which is proved by Theorem \ref{theorem:forward} to hold asymptotically
for the starting set of the backward stage $\tilde{\mathcal{S}}_{F}$.
Hence, combining Theorem \ref{theorem:forward} and \ref{theorem:backward}
establishes the model selection consistency of SODA. Proofs of the theorems are in the on-line Supplemental Materials.

\section{Simulation Results} \label{sec:simstudy}

\subsection{Discriminant analysis with interactions}
We first evaluate performances of a few methods on main and interaction effects selection under the discriminant analysis framework. 
Besides SODA, we consider the backward
procedure in \cite{zhang2011bic} (denoted as ZW), the
forward-backward method in \cite{murphy2010variable}
(denoted as MDR), hierNet in \cite{bien2013lasso}
and IIS-SQDA in \cite{fan2015innovated}. 
Both ZW and MDR require the joint normality between $\mathbf{X}_{\mathcal{P}}$
and $\mathbf{X}_{\mathcal{P}^{c}}$. HierNet is a Lasso-like
procedure to detect multiplicative interactions between predictors
under hierarchical constraints. For hierNet, we select the regularization
parameter with the lowest CV error. We have also reported in the Supplemental Materials a comparison between SODA and Lasso-logistic for variable selections when the underlying model has only linear main effects, and found that SODA was competitive with Lasso in all cases we tested and out-performed Lasso significantly when the ``incoherence"   \citep{ravikumar2010high} or the ``irrepresentable'' \citep{zhao2006model}  condition
was violated.

We first considered four simulation settings in Examples 1.1$\sim$1.4 for the classification example introduced in Section 1 (see Figure 1 for more details), and then examined two more simulation scenarios (Examples 1.5 and 1.6) in which the interaction effects and main effects are from different predictors. For Examples 1.1$\sim$1.4, there are two classes ($K=2$) and $p$ predictors, among which  $X_{1}$, $X_{2}$ and $X_{3}$ are relevant ones,  i.e.,  $\mathcal{P}=\left\{ 1,2,3\right\}$,  and are simulated as multivariate Gaussian conditional on the class label. Other $p-3$ predictors are irrelevant but correlated with the three relevant ones.
The oracle Bayes classification rule for these four examples is to label an observation class
1 if $Q\left(\mathbf{x}\right)>0$, and 0 otherwise, where 
\[
Q\left(\mathbf{x}\right)=1.627+X_{1}-0.6X_{1}^{2}-0.6X_{3}^{2}-0.7X_{1}X_{2}-0.7X_{2}X_3,
\]
indicating that $\mathcal{A}=\left\{ 1,\left(1,1\right),\left(3,3\right),\left(1,2\right),\left(2,3\right)\right\}$, representing one linear effect  and four interaction
effects without the hierarchy restriction.
The setting of Example 1.1 follows the multivariate normal model
while the other three do not.  Examples
1.1$\sim$1.3 are of moderate dimension with $p$=50, and Example 1.4
simulates a high-dimensional scenario with $p$=1000.

For each simulation setting, we generated 100 datasets with 10 different
sample sizes for each class, ranging linearly in log-scale from $100$
to $1000$: $n=100, [100\times10^{1/9}],\dots,1000$.
For SODA, hierNet, and IIS-SQDA, the set of selected predictor variables
is defined as the union of all predictors appearing in the selected linear
and interaction terms. We calculated the average number of false negatives
and false positives for variable selection (VFN and VFP), main effect
term selection (MFN and MFP), and interaction term selection (IFN
and IFP).

\begin{figure}[h]
\noindent \begin{centering}
\includegraphics[scale=0.775]{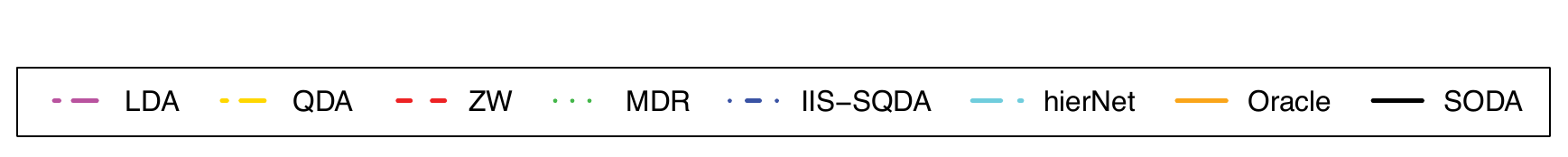}
\par\end{centering}

\noindent \begin{centering}
\includegraphics[scale=1,height=1.7in,width=2.8in]{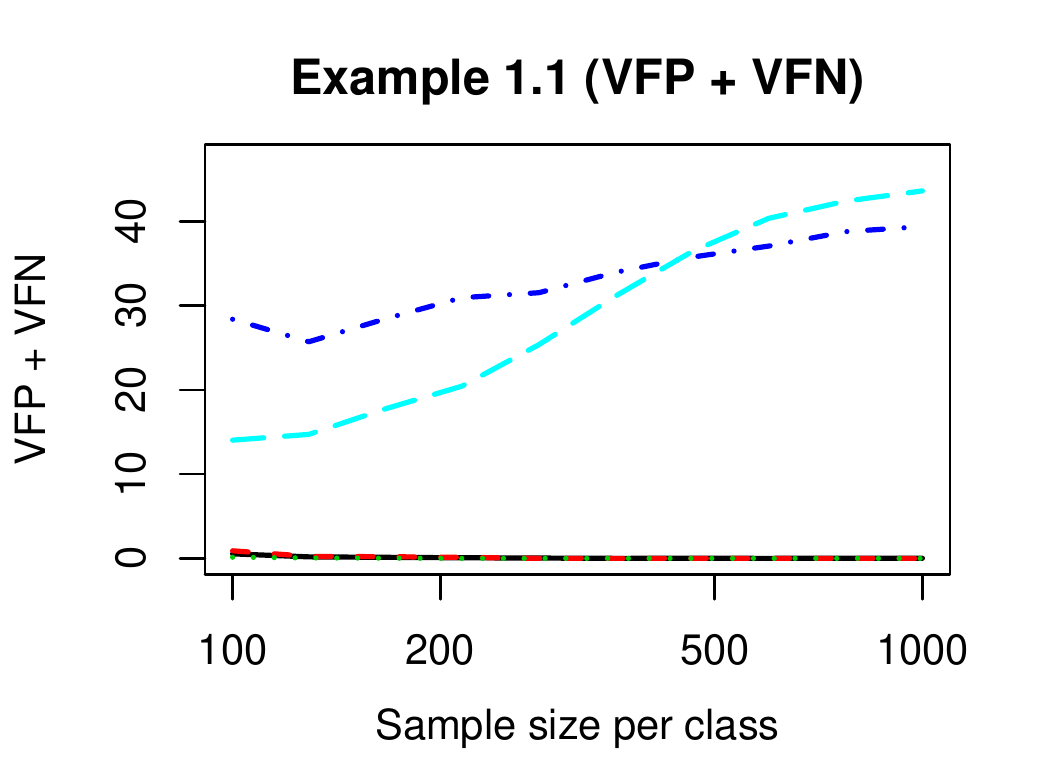} \includegraphics[scale=1,height=1.7in,width=2.8in]{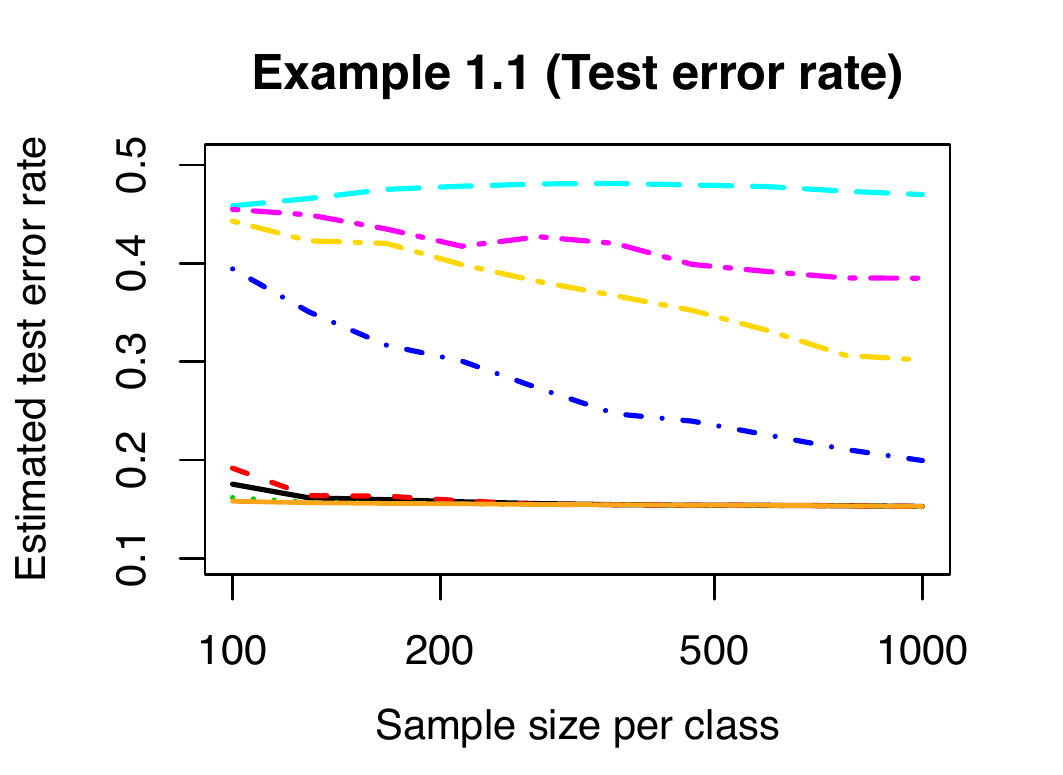}
\par\end{centering}

\noindent \begin{centering}
\includegraphics[scale=1,height=1.7in,width=2.8in]{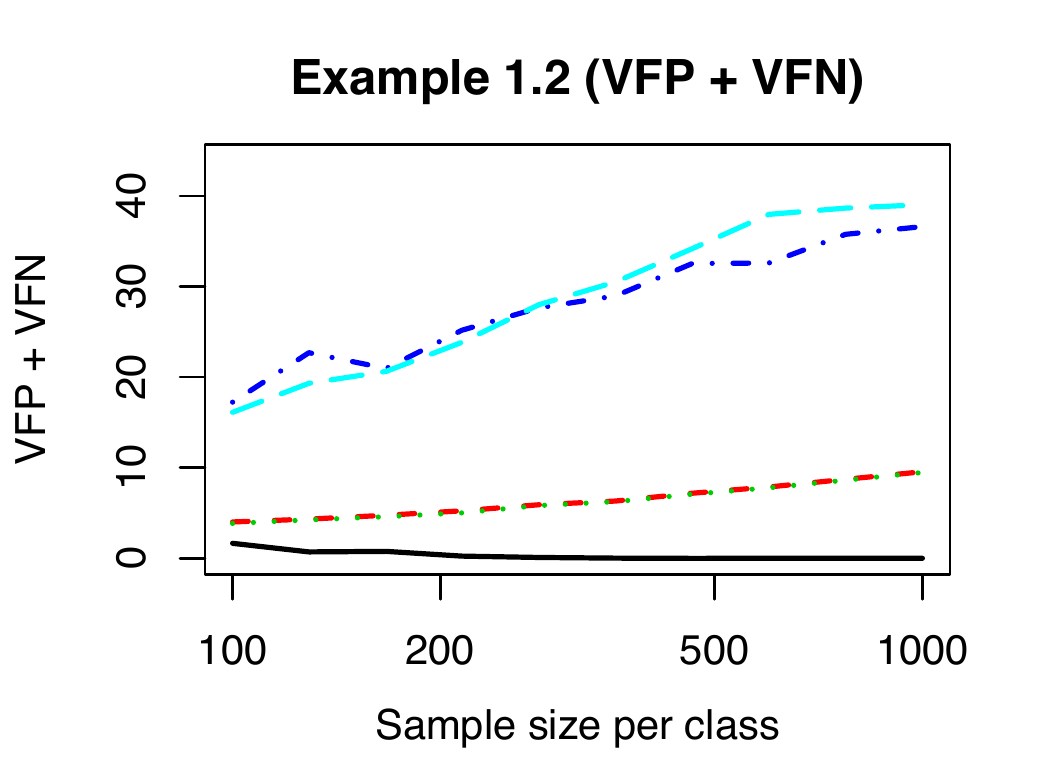} \includegraphics[scale=1,height=1.7in,width=2.8in]{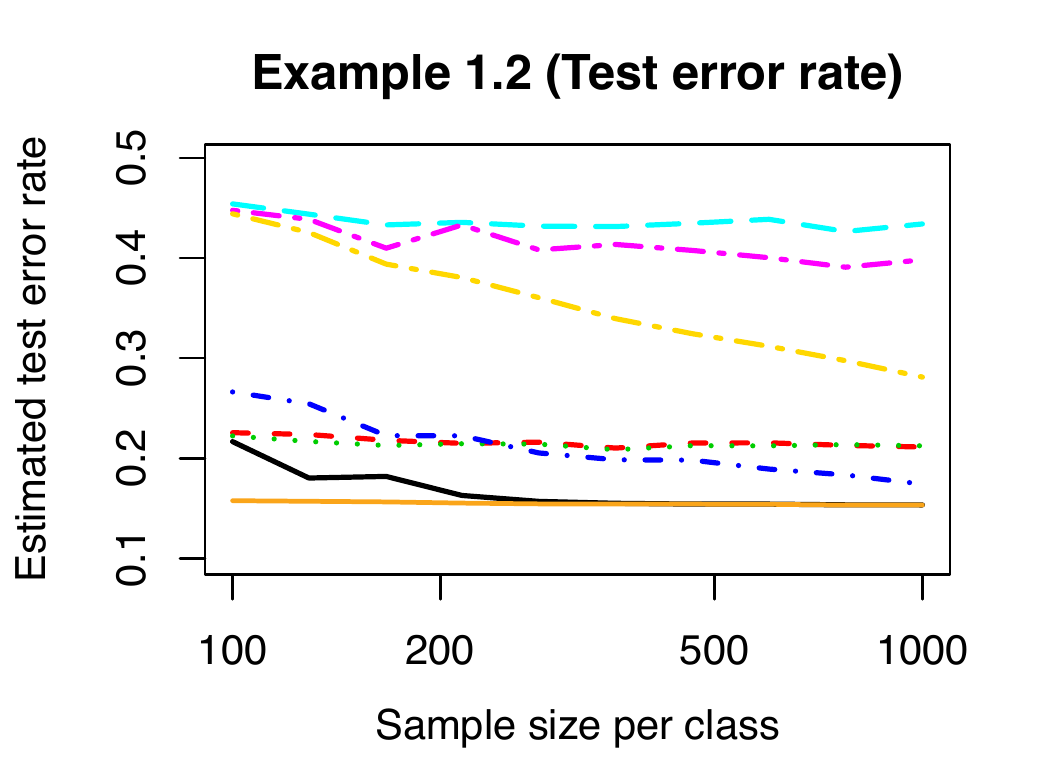}
\par\end{centering}

\noindent \begin{centering}
\includegraphics[scale=1,height=1.7in,width=2.8in]{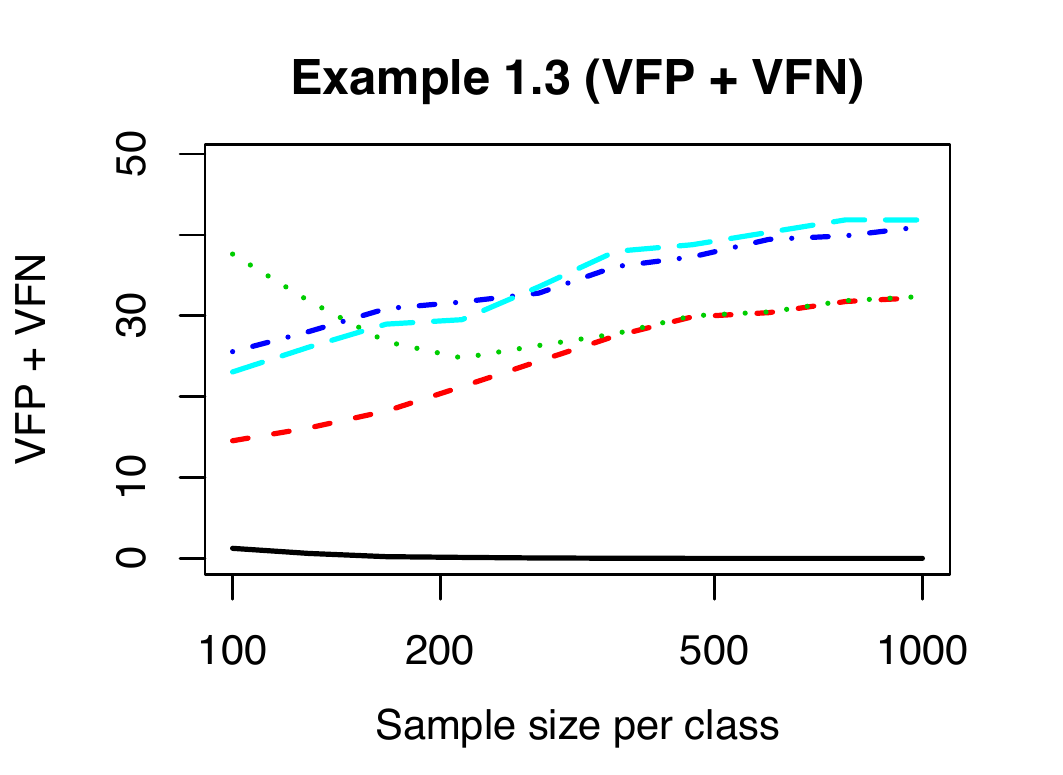} \includegraphics[scale=1,height=1.7in,width=2.8in]{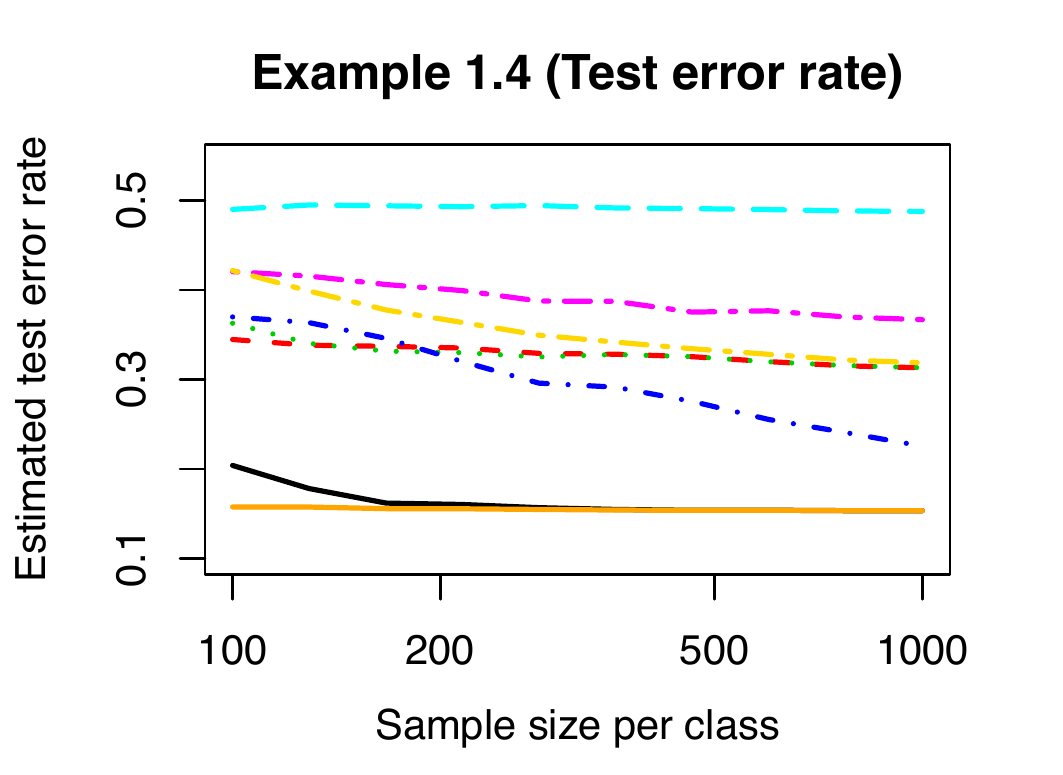}
\par\end{centering}

\noindent \begin{centering}
\includegraphics[scale=1,height=1.7in,width=2.8in]{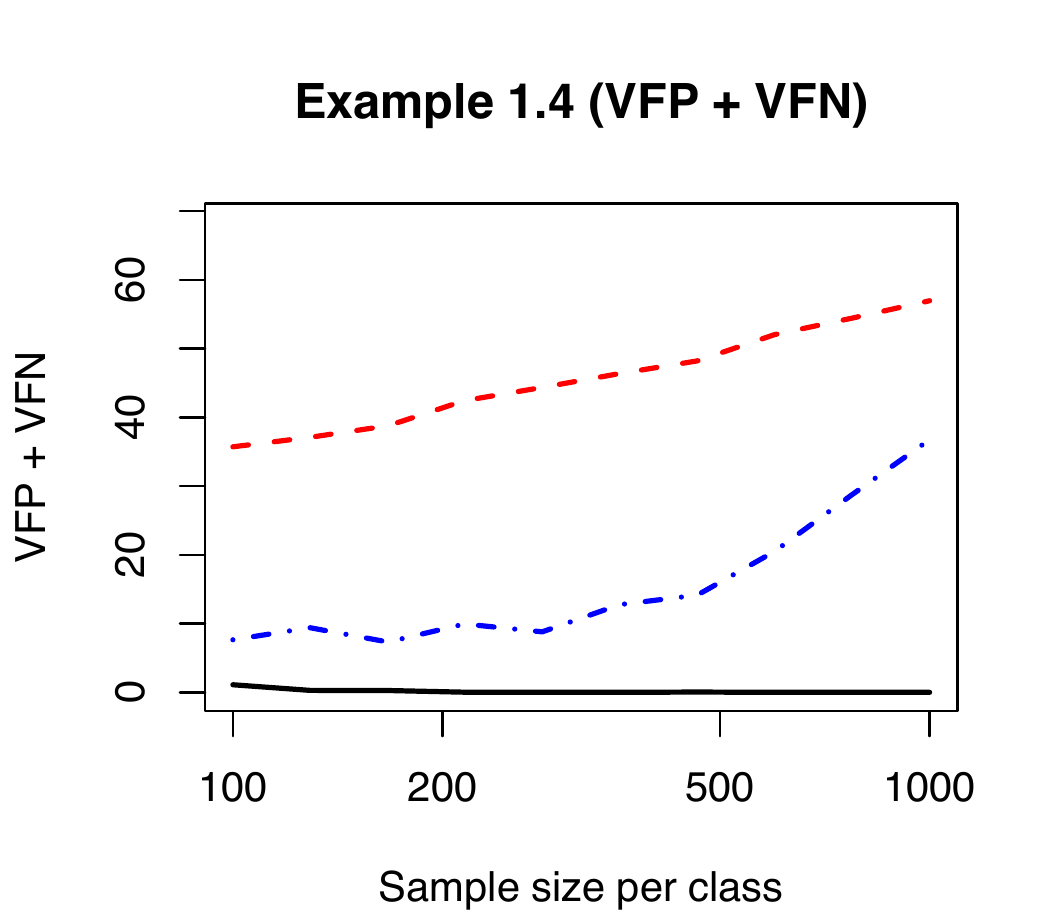} \includegraphics[scale=1,height=1.7in,width=2.8in]{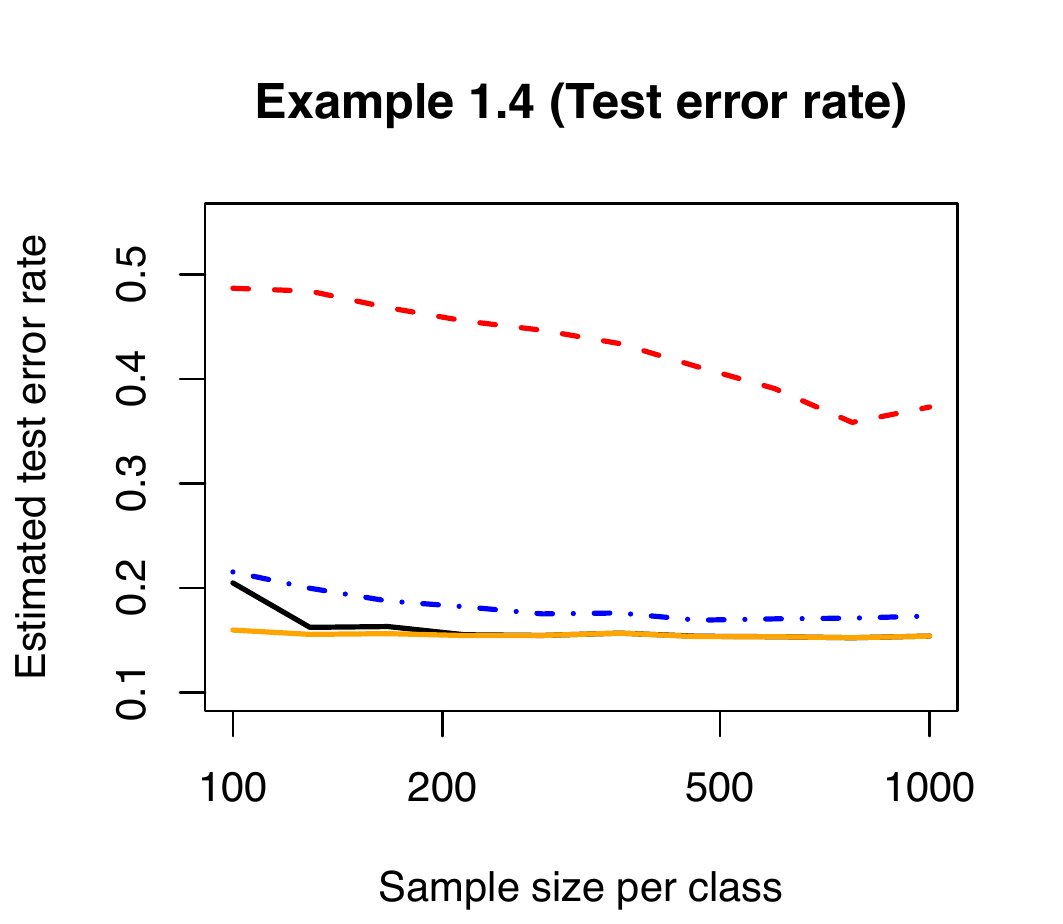}
\par\end{centering}

\caption{Results for Example 1.1$\sim$1.4. VFP: average number of variable
selection false positives. VFN: average number of variable selection
false negatives. LDA and QDA used all the variables without any selection, so they do not appear in the left panel and their TEs were high. MDR and hierNet all broke down for Example 1.4. LDA and QDA also did not work due to large $p$. \label{fig:sim_1.1-1.3}}
\end{figure}

To benchmark the classification accuracy, we also include the full
model of LDA and QDA with all predictors, and the Oracle model that
contains exactly the five true terms. The average classification test
error rate (TE) of each method is estimated by applying the trained model
to 10,000 extra observations simulated from the true model. 
Results for Examples 1.1$\sim$1.4 are shown in Figure \ref{fig:sim_1.1-1.3}. For SODA, hierNet and IIS-SQDA, we also counted the numbers
of FNs and FPs for the selection of main effect and interaction terms, respectively, and show them  in Table \ref{tab:sim_1.1-1.4_MI}.

\vspace{3mm}

\textbf{Example 1.1. } {\it Multivariate Gaussian}. Irrelevant predictors
were simulated as linear combinations of relevant ones as follows:
\[
X_{j}=b_{j,0}+b_{j,1}X_{l}+b_{j,2}X_{k}+\varepsilon_{j}, \ j=3,\ldots, 50,
\]
where $X_{k}$ and $X_{l}$ were randomly selected from $\{ X_{1}, X_{2},
X_{3}\}$, coefficients $b_{j,0}$, $b_{j,1}$ and $b_{j,2}$ were
drawn from uniform distribution $U\left[-1,1\right]$, and $\varepsilon_{j}\sim\mathcal{N}\left(0,2\right)$. 

As shown in Figure~\ref{fig:sim_1.1-1.3}, for this example ZW, MDR, and SODA were all able to detect all relevant predictors as $n$ increases, with both VFN and VFP being very low. They achieved almost the Oracle classification accuracy.  In contrast, IIS-SQDA and hierNet
selected too many false positives, which resulted in high test error rates, and the number VFP+VFN increased with
$n$. This strange phenomenon has also been observed by other researchers \citep{fan2015innovated,yu2014modified}.  
The performances of IIS-SQDA, hierNet and SODA on individual term selection are shown in Table~\ref{tab:sim_1.1-1.4_MI}. 
SODA selected individual terms nearly perfectly. 
HierNet is based on Lasso and IIS-SQDA uses elastic net. The variable selection consistency of Lasso and elastic net require the {\it Irrepresentable Condition} \citep{zhao2006model} and the {\it Elastic Irrepresentable Condition} \citep{jia2010model}, which may not hold here. Moreover, it was observed that the cross-validation is too liberal for Lasso, leading to a large number of false positives \citep{yu2014modified}. As expected, LDA and QDA without variable selection performed the worst.  

\vspace{3mm}

\textbf{Example 1.2. } {\it Non-Gaussian irrelevant predictors}. Irrelevant variables were simulated to be quadratically dependent of relevant ones: 
\begin{equation}
X_{j}=b_{j,0}+b_{j,1}X_{k}+b_{j,2}X_{l}+b_{j,3}X_{k}^{2}+b_{j,4}X_{l}^{2}+\varepsilon_{j}, \ j=3,\ldots, 50, \label{eq:sim_4}
\end{equation}
where $X_{k}$ and $X_{l}$ were randomly selected from $\{ X_{1}, X_{2},
X_{3}\}$, coefficients $b_{j,0}$, ..., $b_{j,4}$ were
drawn from $U\left[-1,1\right]$, and $\varepsilon_{j}\sim\mathcal{N}\left(0,5\right)$.
As shown in Figure \ref{fig:sim_1.1-1.3}, ZW and MDR  selected
 4 to 10 FP and FN predictors on average. IIS-SQDA
and hierNet selected a large number of FP terms,
as shown in Table~\ref{tab:sim_1.1-1.4_MI}, due to the correlation
between relevant and irrelevant predictors as well as correlations between
main and interaction terms.

\vspace{3mm}

\textbf{Example 1.3. }{\it Heteroskedastic covariates.} Irrelevant we simulated 
as follows: 
\begin{equation}
X_{j}=b_{j,1}X_{k}+b_{j,2}X_{l}+\left|X_{k}\right|\varepsilon_{j}, \ j=3,\ldots, 50, \label{eq:sim_5}
\end{equation}
where $X_{k}$ and $X_{l}$ were randomly selected from $\{ X_{1}, X_{2},
X_{3}\}$, coefficients $b_{j,1}$ and $b_{j,3}$ were
drawn from $U\left[-1,1\right]$, and $\varepsilon_{j}\sim\mathcal{N}\left(0,1\right)$.
It violates the constant variance assumption of ZW and MDR. Thus, ZW, MDR, IIS-SQDA and hierNet all performed sub-optimally. In contrast,
SODA selected almost no VFP and VFN, and achieved near-Oracle prediction
accuracy when $n\ge200$.

\vspace{3mm}

\begin{figure}[h]
\noindent \begin{centering}
\includegraphics[height=2.5in,width=4in]{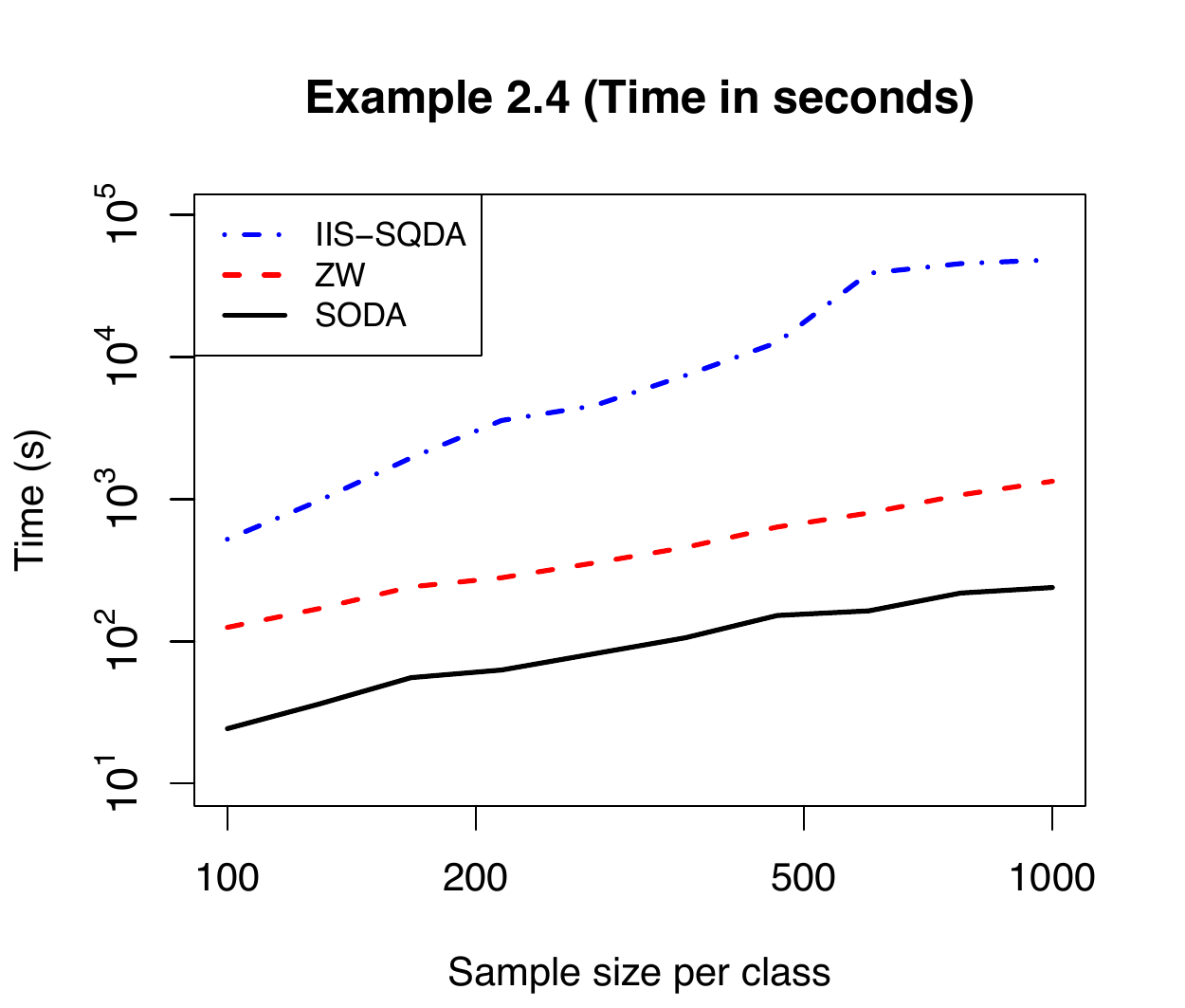} 
\par\end{centering}

\caption{Mean running time in seconds for ZW, IIS-SQDA, and SODA for Example
1.4; and hierNet did not finish the job within 24 hours.\label{fig:sim_1.4_time}}
\end{figure}

\textbf{Example 1.4. }{\it High-dimensional and non-Gaussian.}  Irrelevant
predictors were simulated as follows. For $j\in\left\{ 4,\dots,100\right\} $,
we drew 60\% of the $X_j$'s at random  and simulated them from $\mathcal{N}\left(m_{j},1\right)$,
$m_{j}\sim U\left[0,1\right]$.  The remaining $40\%$ of
the $X_j$'s were simulated as  non-linearly related to $\left(X_{k},X_{l}\right)$ similarly as (\ref{eq:sim_4}) or
(\ref{eq:sim_5}), where $k$ and $l$ were randomly chosen from $\left\{ 1,2,3\right\} $.
For $j\in\left\{ 101,\dots,1000\right\} $, we first drew all predictors
from $\mathcal{N}\left(m_{j},1\right)$, and then randomly selected
$40\%$ of them and re-simulated each of the selected $X_j$  as (\ref{eq:sim_4})
or (\ref{eq:sim_5}), where $k$ and $l$ are indexes uniformly drawn
from $\{101,\ldots, 1000\}$. We changed ZW to a forward procedure
since the backward procedure is not feasible when $p>n$. Results
are shown in Figure \ref{fig:sim_1.1-1.3} and Table \ref{tab:sim_1.1-1.4_MI}.
MDR results are not shown because it is unstable for highly correlated
$\mathbf{X}$ matrices and usually keeps on adding new predictors until
the estimation of covariance matrices become singular. Overall, SODA
performed much better than ZW and IIS-SQDA, and achieved near-oracle 
the classification accuracy for $n>100$. Figure~\ref{fig:sim_1.4_time}
shows the running times in log-scale versus $n$  for IIS-SQDA, ZW, and SODA,
On average, IIS-SQDA took 800 minutes, ZW took 22 minutes, and SODA took 4 minutes to analyze one simulated dataset with $p=1000$ and $n=1000$. In contrast, hierNet did not finish the simulation experiments in 24 hours and is thus not included in the comparison. 

\begin{table}[h]
\begin{centering}
{\footnotesize{}}%
\begin{tabular}{|cc|cccc|cccc|cccc|}
\hline 
 &  & \multicolumn{4}{c|}{\textbf{\footnotesize{}SODA}} & \multicolumn{4}{c|}{\textbf{\footnotesize{}IIS-SQDA}} & \multicolumn{4}{c|}{\textbf{\footnotesize{}hierNet}}\tabularnewline[\doublerulesep]
\cline{3-14} 
{\footnotesize{}Example} & \textbf{\footnotesize{}$n$} & \textbf{\footnotesize{}MFN} & \textbf{\footnotesize{}MFP} & \textbf{\footnotesize{}IFN} & \textbf{\footnotesize{}IFP} & \textbf{\footnotesize{}MFN} & \textbf{\footnotesize{}MFP} & \textbf{\footnotesize{}IFN} & \textbf{\footnotesize{}IFP} & \textbf{\footnotesize{}MFN} & \textbf{\footnotesize{}MFP} & \textbf{\footnotesize{}IFN} & \textbf{\footnotesize{}IFP}\tabularnewline[\doublerulesep]
\hline 
 & {\footnotesize{}$100$} & \textbf{\footnotesize{}$0.05$} & {\footnotesize{}$0.16$} & {\footnotesize{}$1.01$} & {\footnotesize{}$0.30$} & {\footnotesize{}$0.27$} & {\footnotesize{}$2.39$} & {\footnotesize{}$0.90$} & {\footnotesize{}$48.5$} & {\footnotesize{}$0$} & {\footnotesize{}$12.6$} & {\footnotesize{}$1.58$} & {\footnotesize{}$2.26$}\tabularnewline[\doublerulesep]
{\footnotesize{}1.1} & {\footnotesize{}$215$} & {\footnotesize{}$0$} & {\footnotesize{}$0.01$} & {\footnotesize{}$0.04$} & {\footnotesize{}$0.02$} & {\footnotesize{}$0.08$} & {\footnotesize{}$2.90$} & {\footnotesize{}$0.25$} & {\footnotesize{}$63.2$} & {\footnotesize{}$0$} & {\footnotesize{}$19.2$} & {\footnotesize{}$1.10$} & {\footnotesize{}$14.0$}\tabularnewline[\doublerulesep]
 & {\footnotesize{}$1000$} & {\footnotesize{}$0$} & {\footnotesize{}$0$} & {\footnotesize{}$0$} & {\footnotesize{}$0$} & {\footnotesize{}$0$} & {\footnotesize{}$6.39$} & {\footnotesize{}$0$} & {\footnotesize{}$112$} & {\footnotesize{}$0$} & {\footnotesize{}$44.6$} & {\footnotesize{}$0$} & {\footnotesize{}$46.2$}\tabularnewline[\doublerulesep]
\hline 
 & {\footnotesize{}$100$} & {\footnotesize{}$0.26$} & {\footnotesize{}$0.58$} & {\footnotesize{}$1.74$} & {\footnotesize{}$0.28$} & {\footnotesize{}$0.26$} & {\footnotesize{}$12.9$} & {\footnotesize{}$0.40$} & {\footnotesize{}$7.42$} & {\footnotesize{}$0$} & {\footnotesize{}$14.9$} & {\footnotesize{}$1.44$} & {\footnotesize{}$7.24$}\tabularnewline[\doublerulesep]
{\footnotesize{}1.2} & {\footnotesize{}$215$} & {\footnotesize{}$0$} & {\footnotesize{}$0.13$} & {\footnotesize{}$0.27$} & {\footnotesize{}$0.03$} & {\footnotesize{}$0$} & {\footnotesize{}$19.7$} & {\footnotesize{}$0.02$} & {\footnotesize{}$11.1$} & {\footnotesize{}$0$} & {\footnotesize{}$22.9$} & {\footnotesize{}$0.65$} & {\footnotesize{}$15.3$}\tabularnewline[\doublerulesep]
 & {\footnotesize{}$1000$} & {\footnotesize{}$0$} & {\footnotesize{}$0$} & {\footnotesize{}$0$} & {\footnotesize{}$0$} & {\footnotesize{}$0$} & {\footnotesize{}$28.5$} & {\footnotesize{}$0$} & {\footnotesize{}$24.3$} & {\footnotesize{}$0$} & {\footnotesize{}$39.1$} & {\footnotesize{}$0$} & {\footnotesize{}$46.9$}\tabularnewline[\doublerulesep]
\hline 
 & {\footnotesize{}$100$} & {\footnotesize{}$0.12$} & {\footnotesize{}$0.13$} & {\footnotesize{}$1.50$} & {\footnotesize{}$0.70$} & {\footnotesize{}$0.09$} & {\footnotesize{}$5.59$} & {\footnotesize{}$0.13$} & {\footnotesize{}$44.9$} & {\footnotesize{}$0.04$} & {\footnotesize{}$21.4$} & {\footnotesize{}$2.20$} & {\footnotesize{}$19.3$}\tabularnewline[\doublerulesep]
{\footnotesize{}1.3} & {\footnotesize{}$215$} & {\footnotesize{}$0.02$} & {\footnotesize{}$0.03$} & {\footnotesize{}$0.17$} & {\footnotesize{}$0.07$} & {\footnotesize{}$0$} & {\footnotesize{}$8.96$} & {\footnotesize{}$0$} & {\footnotesize{}$61.1$} & {\footnotesize{}$0$} & {\footnotesize{}$29.5$} & {\footnotesize{}$0.93$} & {\footnotesize{}$25.7$}\tabularnewline[\doublerulesep]
 & {\footnotesize{}$1000$} & {\footnotesize{}$0$} & {\footnotesize{}$0$} & {\footnotesize{}$0$} & {\footnotesize{}$0$} & {\footnotesize{}$0$} & {\footnotesize{}$14.71$} & {\footnotesize{}$0$} & {\footnotesize{}$99.8$} & {\footnotesize{}$0$} & {\footnotesize{}$42.8$} & {\footnotesize{}$0$} & {\footnotesize{}$46.0$}\tabularnewline[\doublerulesep]
\hline 
 & {\footnotesize{}$100$} & {\footnotesize{}$0.20$} & {\footnotesize{}$0.22$} & {\footnotesize{}$1.58$} & {\footnotesize{}$0.30$} & {\footnotesize{}$0.68$} & {\footnotesize{}$1.58$} & {\footnotesize{}$0.42$} & {\footnotesize{}$6.08$} & \textbf{\footnotesize{} } & \textbf{\footnotesize{} } & \textbf{\footnotesize{} } & \textbf{\footnotesize{} }\tabularnewline[\doublerulesep]
{\footnotesize{}1.4} & {\footnotesize{}$215$} & {\footnotesize{}$0$} & {\footnotesize{}$0$} & {\footnotesize{}$0.14$} & {\footnotesize{}$0$} & {\footnotesize{}$0.20$} & {\footnotesize{}$1.74$} & {\footnotesize{}$0.10$} & {\footnotesize{}$8.84$} & \textbf{\footnotesize{} } & \textbf{\footnotesize{} } & \textbf{\footnotesize{} } & \textbf{\footnotesize{} }\tabularnewline[\doublerulesep]
 & {\footnotesize{}$1000$} & {\footnotesize{}$0$} & {\footnotesize{}$0$} & {\footnotesize{}$0$} & {\footnotesize{}$0$} & {\footnotesize{}$0$} & {\footnotesize{}$3.68$} & {\footnotesize{}$0$} & {\footnotesize{}$40.7$} & \textbf{\footnotesize{} } & \textbf{\footnotesize{} } & \textbf{\footnotesize{} } & \textbf{\footnotesize{} }\tabularnewline[\doublerulesep]
\hline 
\end{tabular}
\par\end{centering}{\footnotesize \par}

\caption{Variable Selection Results for Examples 1.1 $\sim$ 1.4.  MFP / MFN: Average number of main effect false positives and negatives. IFP / IFN: Average number of interaction effect false positives and negatives. 
The  number of observations for each class is denoted by $n$.
\label{tab:sim_1.1-1.4_MI}}
\end{table}




\vspace{3mm}

\textbf{Example 1.5. }{\it Interactions only.} We simulated the scenario
in which there are only interaction effects. In particular, we removed the
main effect term $X_{1}$ from the previous classification rule so that
the new classification function becomes
\[
Q\left(\mathbf{x}\right)=1.777-0.6X_{1}^{2}-0.6X_{3}^{2}-0.7X_{1}X_{2}-0.7X_{2}X_{3}.
\]

\vspace{3mm}

\textbf{Example 1.6. }{\it Anti-hierarchical interactions.} We adopt the
terminology ``anti-hierarchical'' from \cite{bien2013lasso}, which
refers to the scenario that the main effects and interaction effects
are from different set of predictors. In this example, the classification function  $Q\left(\mathbf{x}\right)$ is
\[
Q\left(\mathbf{x}\right)=1.777+X_{4}-X_{5}-0.6X_{1}^{2}-0.6X_{3}^{2}-0.7X_{1}X_{2}-0.7X_{2}X_{3}.
\]

\begin{figure}[h]
\noindent \begin{centering}
\includegraphics[scale=0.775]{sim_legend}
\par\end{centering}

\noindent \begin{centering}
\includegraphics[scale=1,height=1.8in,width=2.8in]{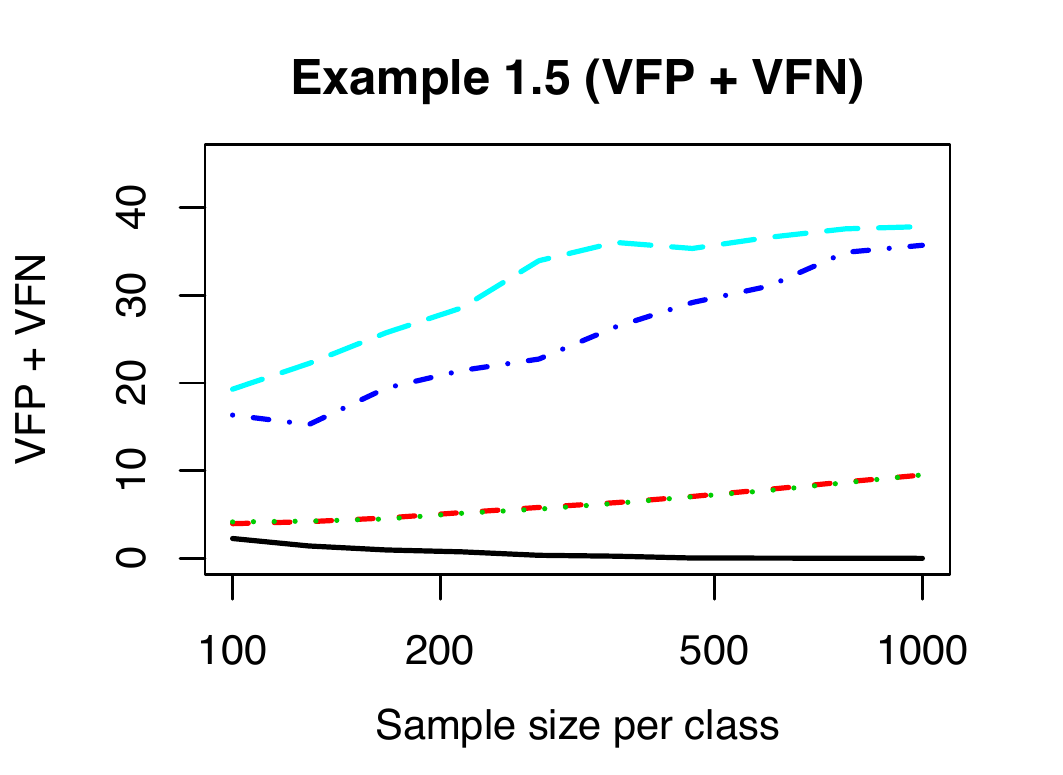}
\includegraphics[scale=1,height=1.8in,width=2.8in]{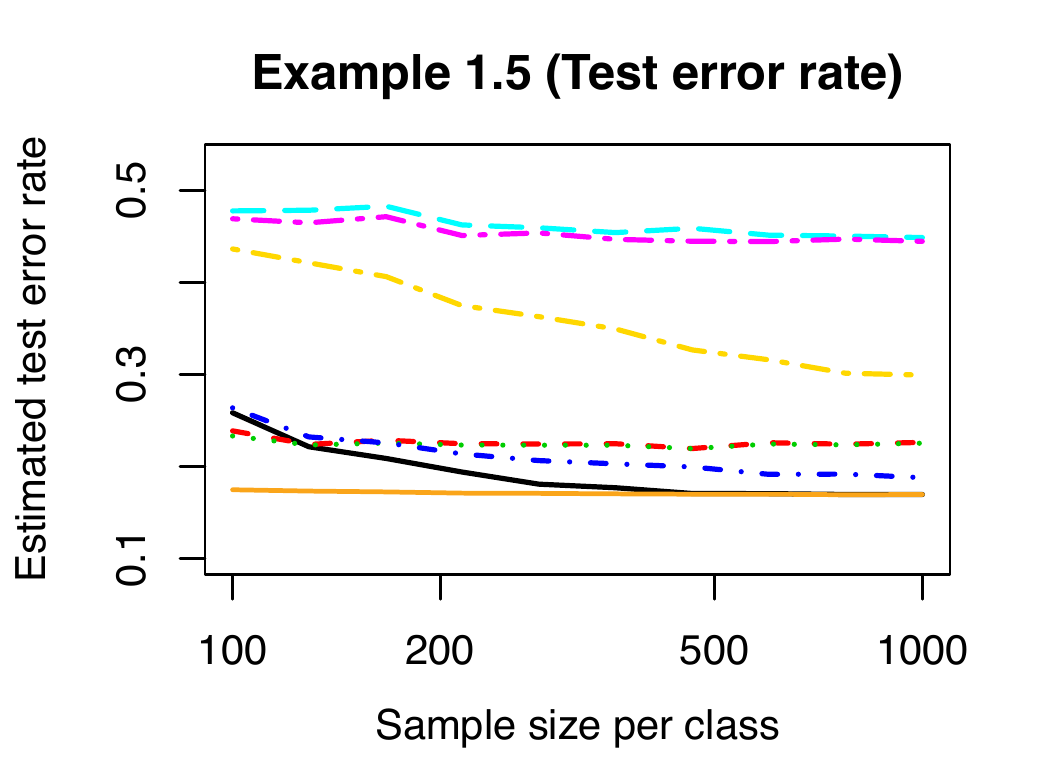}
\par\end{centering}

\noindent \begin{centering}
\includegraphics[scale=1,height=1.8in,width=2.8in]{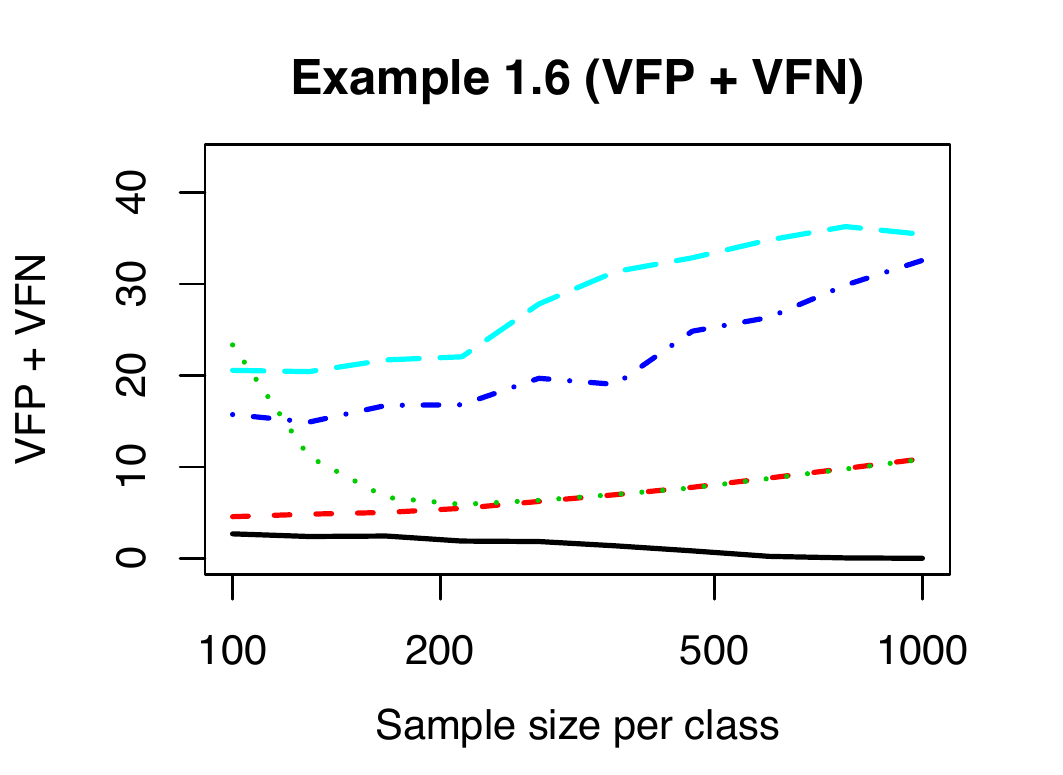}
\includegraphics[scale=1,height=1.8in,width=2.8in]{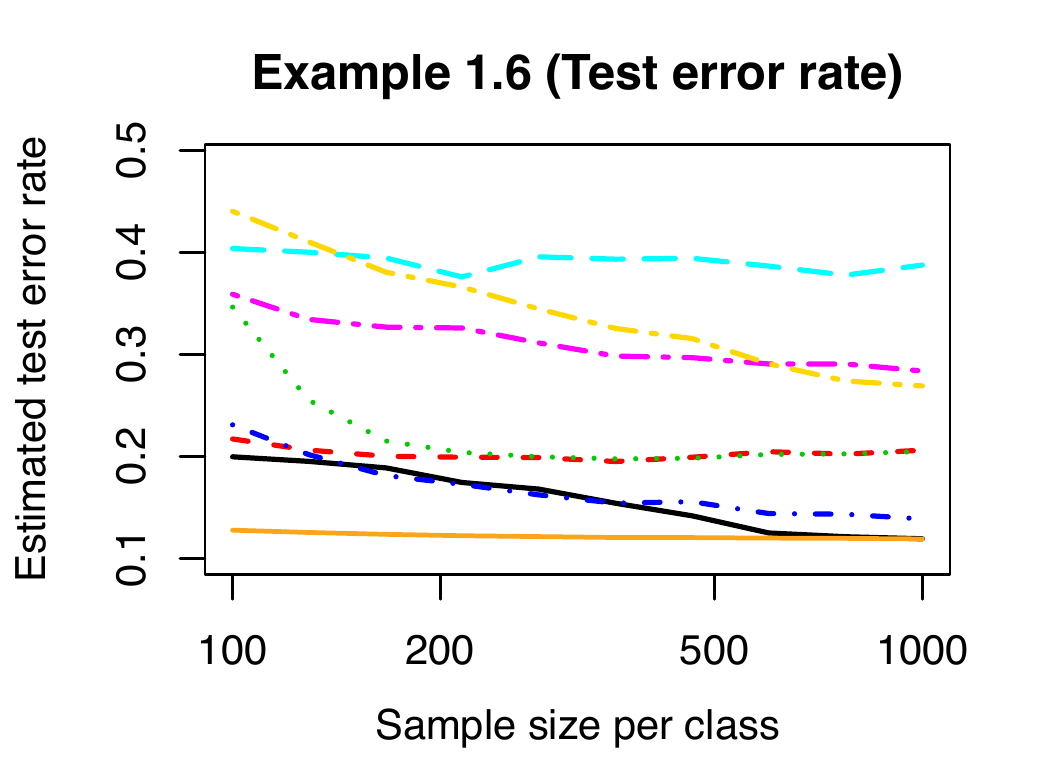}
\par\end{centering}

\caption{Results for Example 1.5 $\sim$ 1.6. VFP: average number of variable
selection false positives. VFN: average number of variable selection
false negatives. \label{fig:sim_1.5-1.6}}
\end{figure}

For both Examples 1.5 and 1.6, we let $p=50$ and let  irrelevant predictors be simulated
in the same way as Example 1.2. The results are shown in Figure \ref{fig:sim_1.5-1.6}. Overall, the results are similar to previous examples that SODA had fewer VFPs and VFNs, and also lower TE rates compared with other methods. In all cases we found that ZW and MDR performed very similarly when $n$ is large.

\subsection{Continuous-response index models}
 
 We examine here  variable selection methods for nonlinear models with continuous responses. Besides S-SODA, we considered all the five methods studied in \cite{jiang2014variable}:
Lasso, DC-SIS, hierNet, COP, and SIRI. DC-SIS  \citep{li2012feature} 
is a sure independence screening procedure based on distance correlation, which has been shown to be capable of detecting relevant variables when interactions are present. HierNet  \citep{bien2013lasso}
is a Lasso-like procedure to detect multiplicative interactions between
predictors under hierarchical constraints.  For SIRI and S-SODA, we
equally partition $\left\{ y_{i}\right\} _{i=1}^{n}$ into $H=5$
slices. In order to improve SIRI's robustness, we consider a modified
version of SIRI, termed as N-SIRI, which pre-processes $\mathbf{X}$
by marginally quantile-normalizing each predictor to the standard
normal distribution.

We considered the following five simulation examples:
\begin{eqnarray*}
\text{Example 2.1}:\qquad Y & = & 3X_{1}+1.5X_{2}+2X_{3}+2X_{4}+2X_{5}+\sigma\epsilon,\\
\text{Example 2.2}:\qquad Y & = & X_{1}+X_{1}X_{2}+X_{1}X_{3}+\sigma\epsilon,\\
\text{Example 2.3}:\qquad Y & = & X_{1}^{2}X_{2}/X_{3}^{2}+\sigma\epsilon,\\
\text{Example 2.4}:\qquad Y & = & X_{1}/\exp\left(X_{2}+X_{3}\right)+\sigma\epsilon,\\
\text{Example 2.5}:\qquad Y & = & X_{1}+X_{2}+\left(1+X_{3}\right)^{2}\epsilon,
\end{eqnarray*}
where $\sigma=0.2$ and $\varepsilon\sim N(0,1)$  independent of $\mathbf{X}$. 
In each example, we simulated the predictors $\mathbf{X}$ with dimension
$p=1000$. In order to test robustness of the methods, we simulated $\mathbf{X}$  under three scenarios.
\begin{itemize}
\item \vspace{-2pt}Scenario (a): $\mathbf{X}$ is simulated from multivariate
Gaussian with correlation $0.5^{\left|i-j\right|}$. In this scenario
the linearity and constant variance conditions hold.
\item \vspace{-5pt}Scenario (b): Each predictor $X_{j}$, $j=1,\dots,p$,
is simulated from the $\chi_{1}^{2}$ distribution independently. In this
scenario the linearity and constant variance conditions
hold, but the distribution of $\mathbf{X}$ is non-normal.
\item \vspace{-5pt}Scenario (c): $X_{1},\dots,X_{125}$ were simulated from
multivariate Gaussian with correlation $0.5^{\left|i-j\right|}$. For $X_{126},\dots,X_{1000}$, we simulated according to the following schemes:
\begin{eqnarray*}
X_{j} & = & X_{j-125}^{2}+\varepsilon_{j},\;\;j=126,\dots,250,\\
X_{j} & = & \sqrt{\left|X_{j-250}\right|}+\varepsilon_{j},\;\;j=251,\dots,375,\\
X_{j} & = & \sin\left(X_{j-375}\right)+\varepsilon_{j},\;\;j=376,\dots,500,\\
X_{j} & = & \log\left(\left|X_{j-500}\right|\right)+\varepsilon_{j},\;\;j=501,\dots,625,\\
X_{j} & = & \exp\left(X_{j-625}\right)+\varepsilon_{j},\;\;j=626,\dots,750,\\
X_{j} & = & \exp\left(\left|X_{j-750}\right|\right)+\varepsilon_{j},\;\;j=751,\dots,875,\\
X_{j} & = & X_{j-875}^{2}\varepsilon_{j},\;\;j=876,\dots,1000.
\end{eqnarray*}
\end{itemize}

\begin{figure}[h]
\begin{centering}
\frame{\includegraphics[height=1.5in, width=4in]{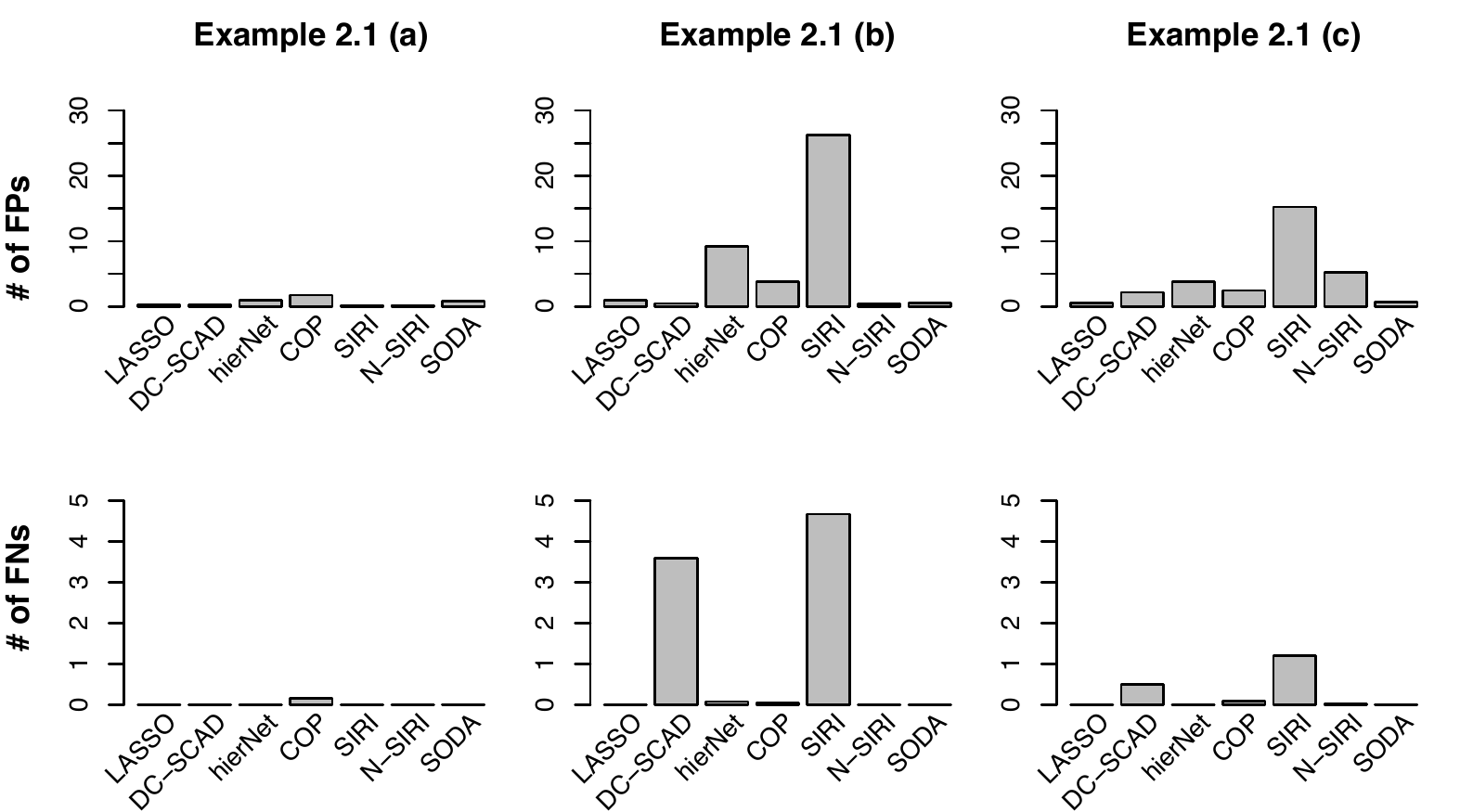}}
\frame{\includegraphics[height=1.5in, width=4in]{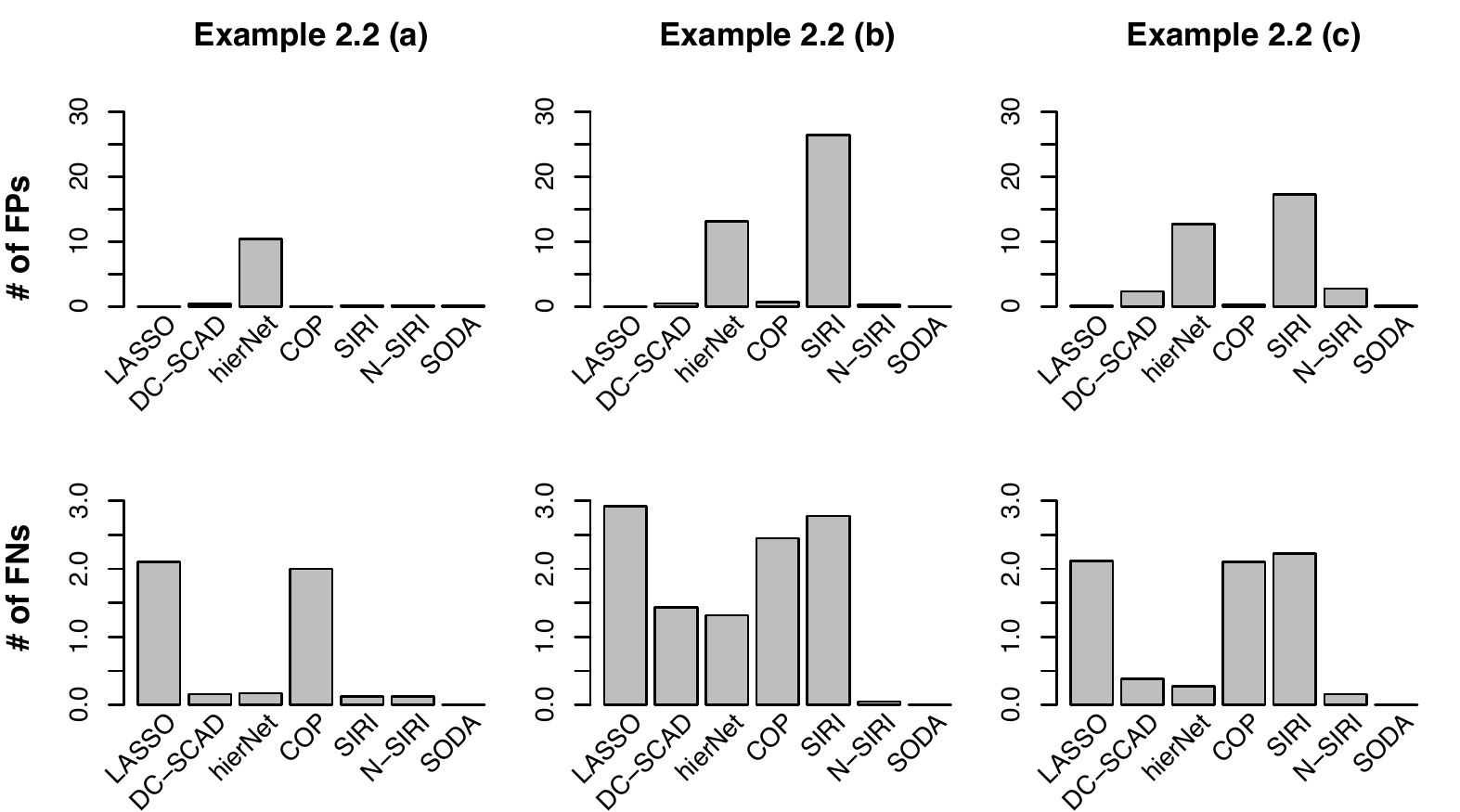}}
\frame{\includegraphics[height=1.5in, width=4in]{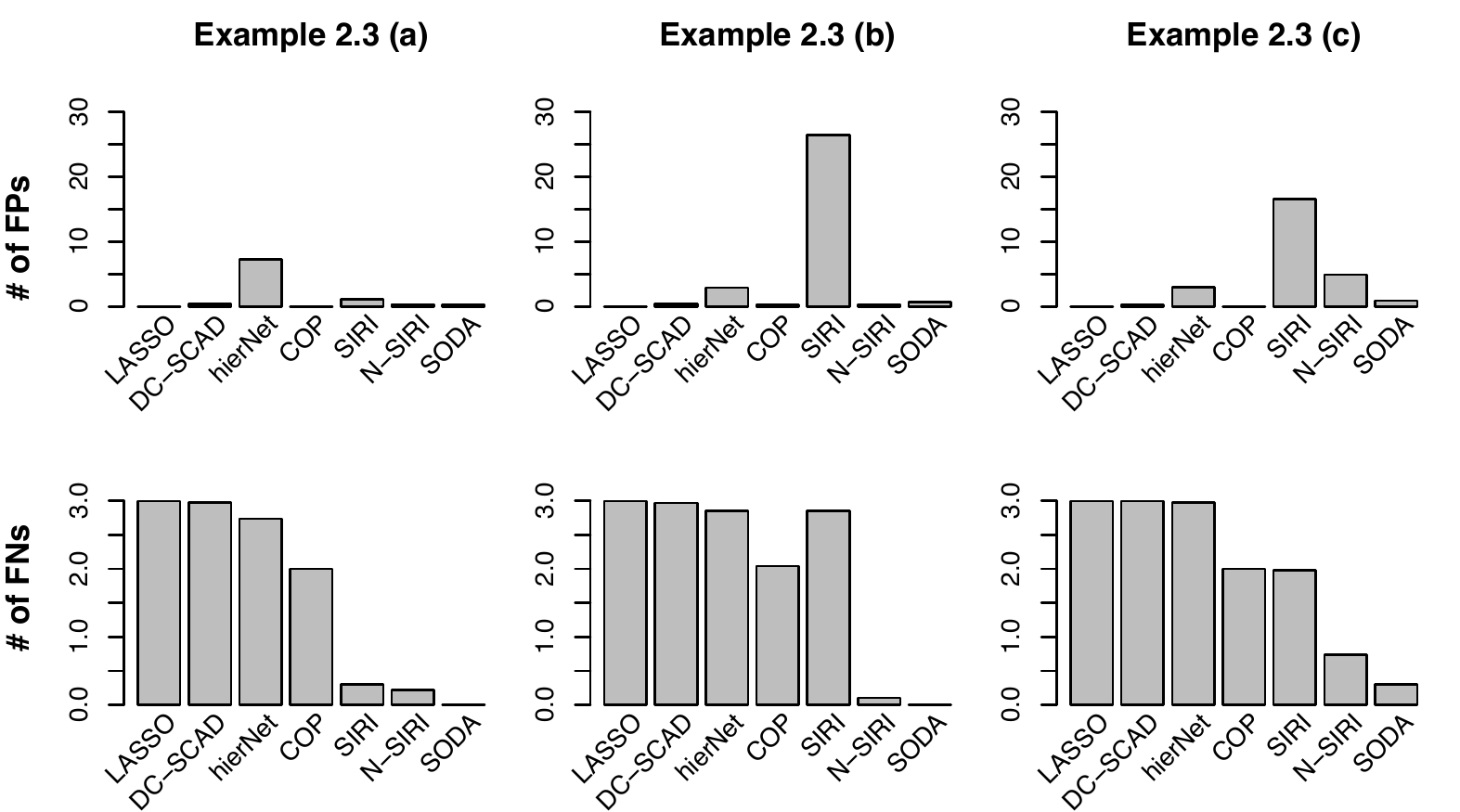}}
\frame{\includegraphics[height=1.5in, width=4in]{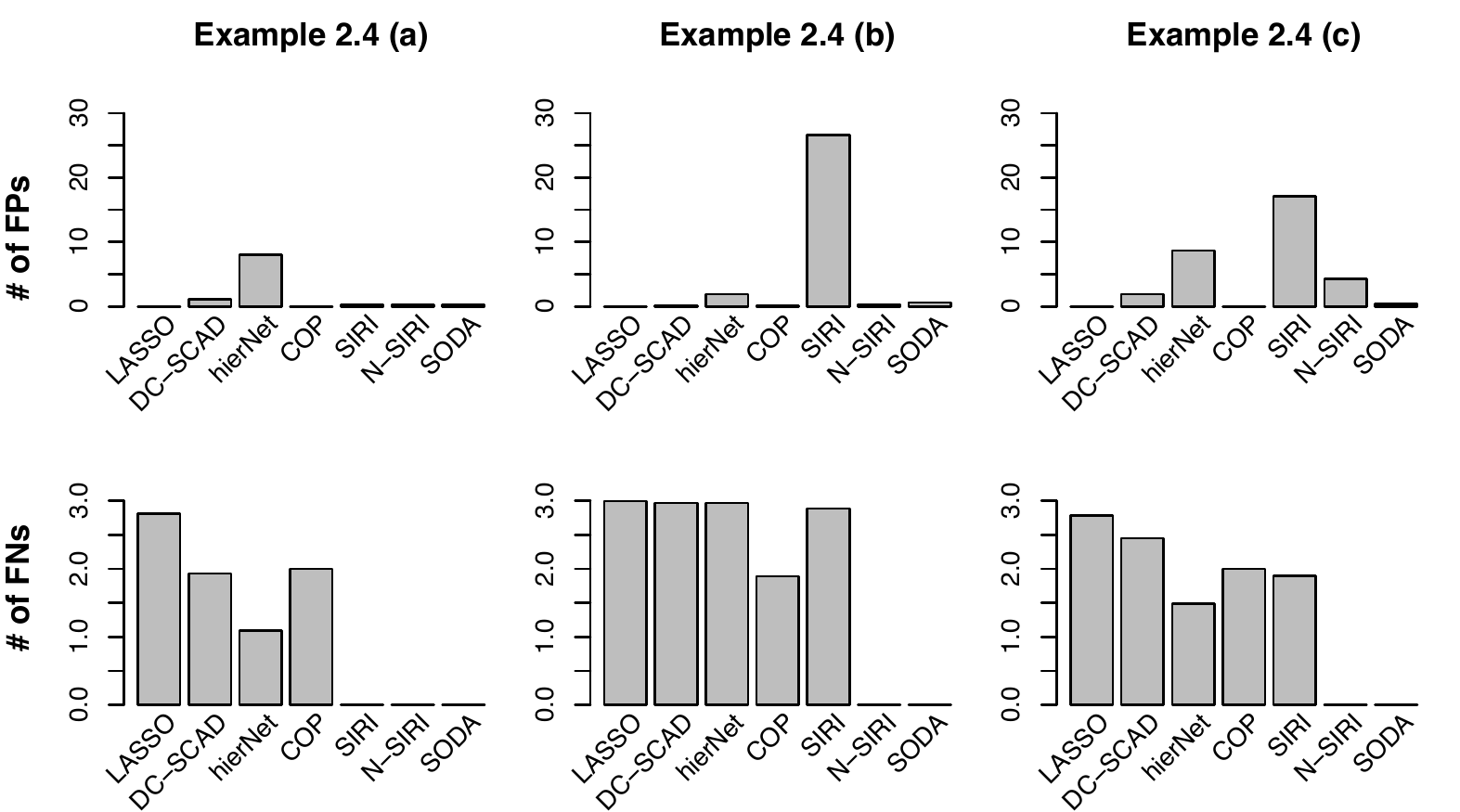}}
\frame{\includegraphics[height=1.5in, width=4in]{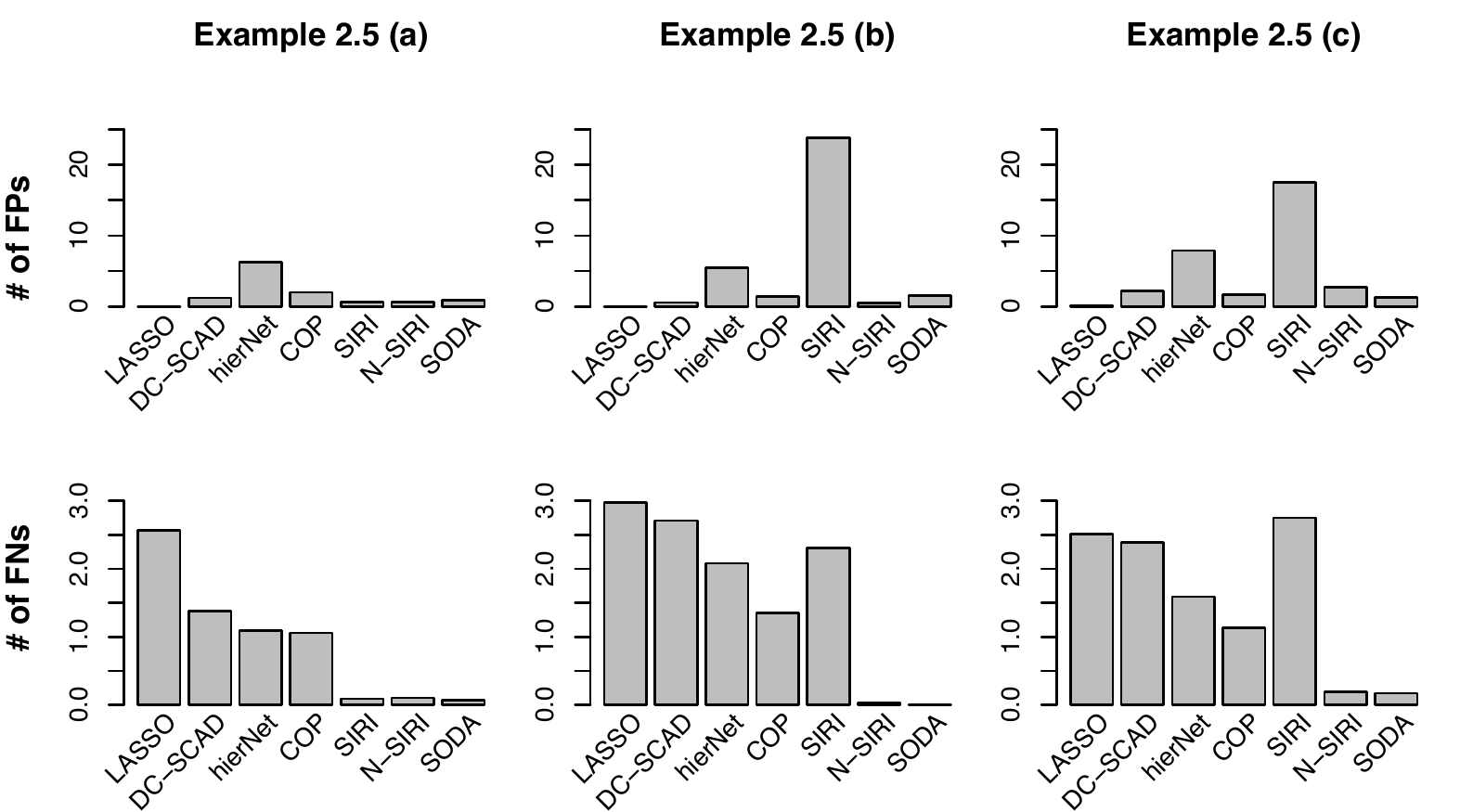}} 
\par\end{centering}
\caption{Simulation study results for Examples 2.1$\sim$2.5. \label{fig:E2}}
\end{figure}

For each simulation setting, we generated 100 datasets with sample
size $n=200$, and applied the aforementioned seven methods to each simulated dataset. For each method, the average number of false positives (FPs) and false
negatives (FNs) were calculated over the 100 datasets. The results
for the five examples are shown in Figure~\ref{fig:E2}. 

As expected, all the seven methods worked well for Example 2.1 in scenario (a), with low FPs and FNs, since the underlying structure is indeed linear Gaussian.  For scenarios (b) and (c) with non-Gaussian predictors,
DC-SCAD, hierNet, and SIRI generated more FPs and/or FNs than other methods. In general. SIRI performed the worst for this example. But with quantile-normalization, N-SIRI performed very competitively. S-SODA worked well for
all the three scenarios, almost as good as Lasso.


In Examples 2.2$\sim$2.5 the relationships between $Y$
and $\mathbf{X}$ is non-linear. Thus, as expected Lasso and DC-SCAD tended to miss important predictors, resulting in high number of FNs. HierNet can only detect second-order interactions, such as $X_{1}X_{2}$, but fails to identify more complicated relationships
such as $Y=X_{1}^{2}X_{2}/X_{3}^{2}$ and $Y=X_{1}/\exp\left(X_{2}+X_{3}\right)$.
COP only identifies the information from the first conditional moment
$\mathbb{E}\left(\mathbf{X}\mid Y\right)$, and misses important
variables with interaction or other second-order effects. 

As expected, SIRI usually worked well for scenario (a). N-SIRI
worked well for both scenarios (a) and (b) since the joint distribution of the predictors become multivariate Gaussian after quantile-normalization.  For scenario (c), SIRI performed very poorly, while N-SIRI performed very respectfully, although  it still had more FPs and FNs than S-SODA. In contrast, S-SODA worked well for all three scenarios. All these examples demonstrated the efficiency and robustness of S-SODA for variable selection in semi-parametric nonlinear regression models.


\subsection{Prediction of continuous surface}
\begin{figure}
\begin{centering}
\includegraphics[viewport=0bp 35bp 576bp 250bp,width=5.5in,height=1.8in]{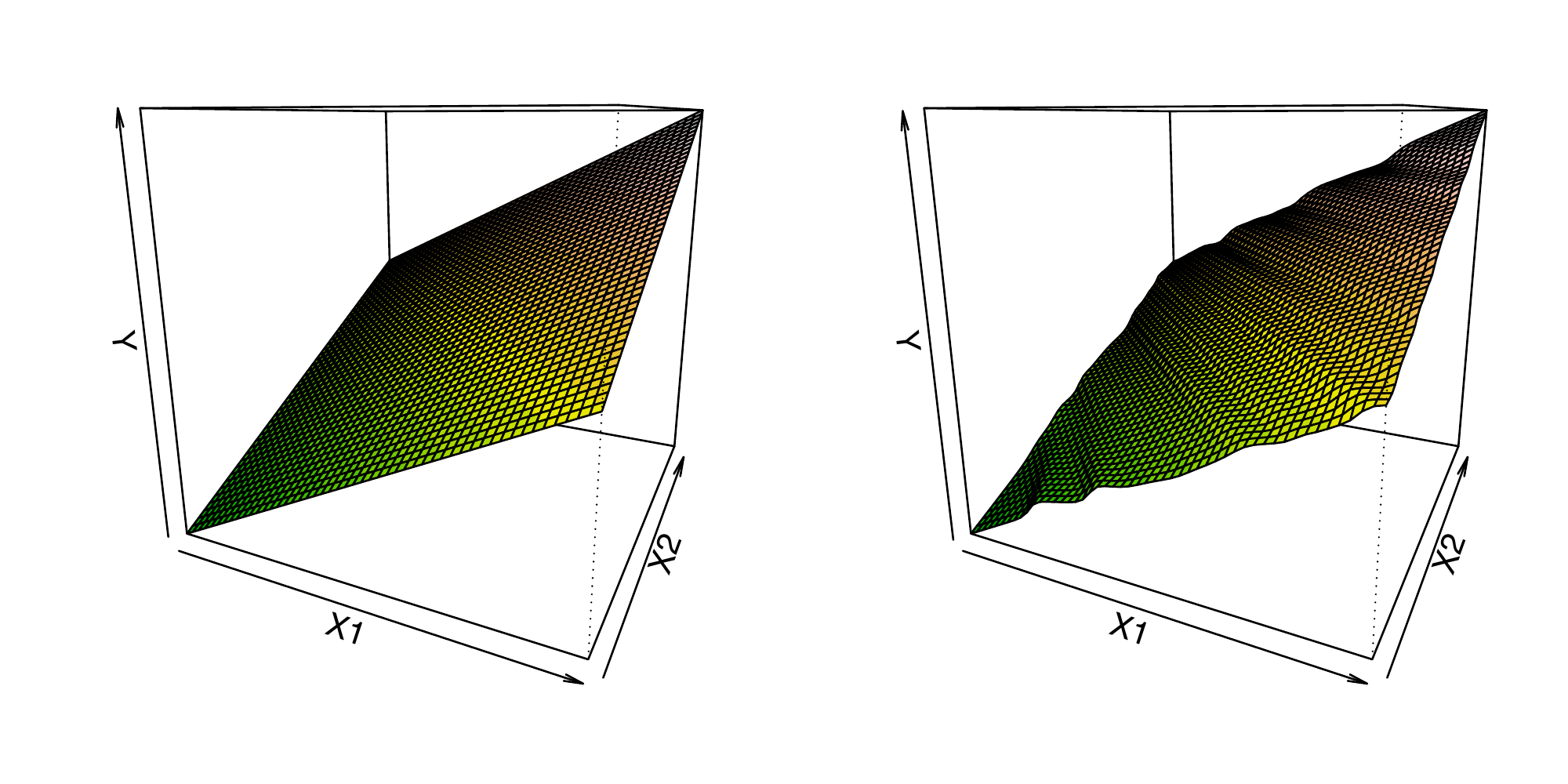}
\includegraphics[viewport=0bp 35bp 576bp 290bp,width=5.5in,height=2.0in]{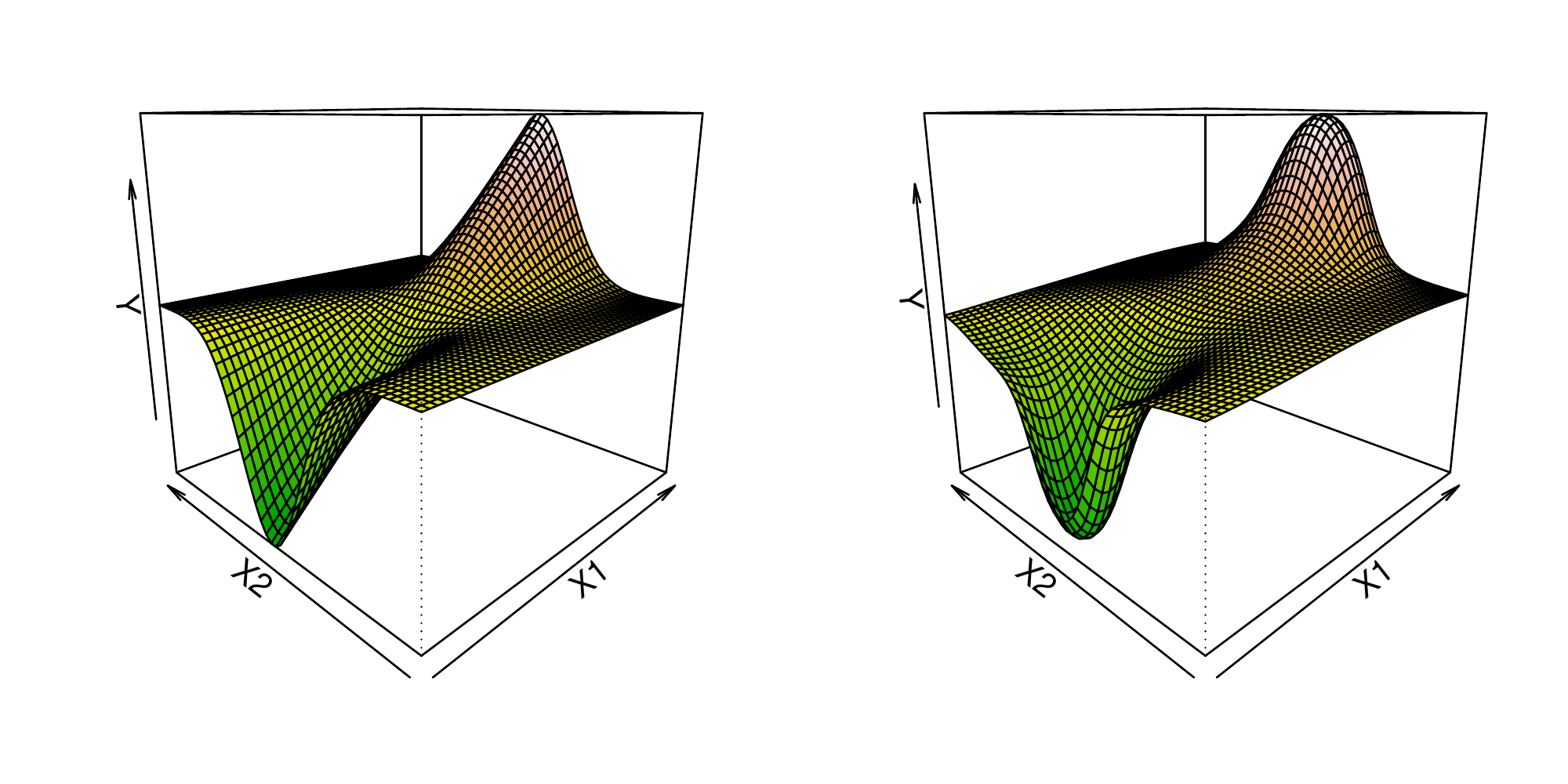}
\includegraphics[viewport=0bp 35bp 576bp 290bp,width=5.5in,height=2.0in]{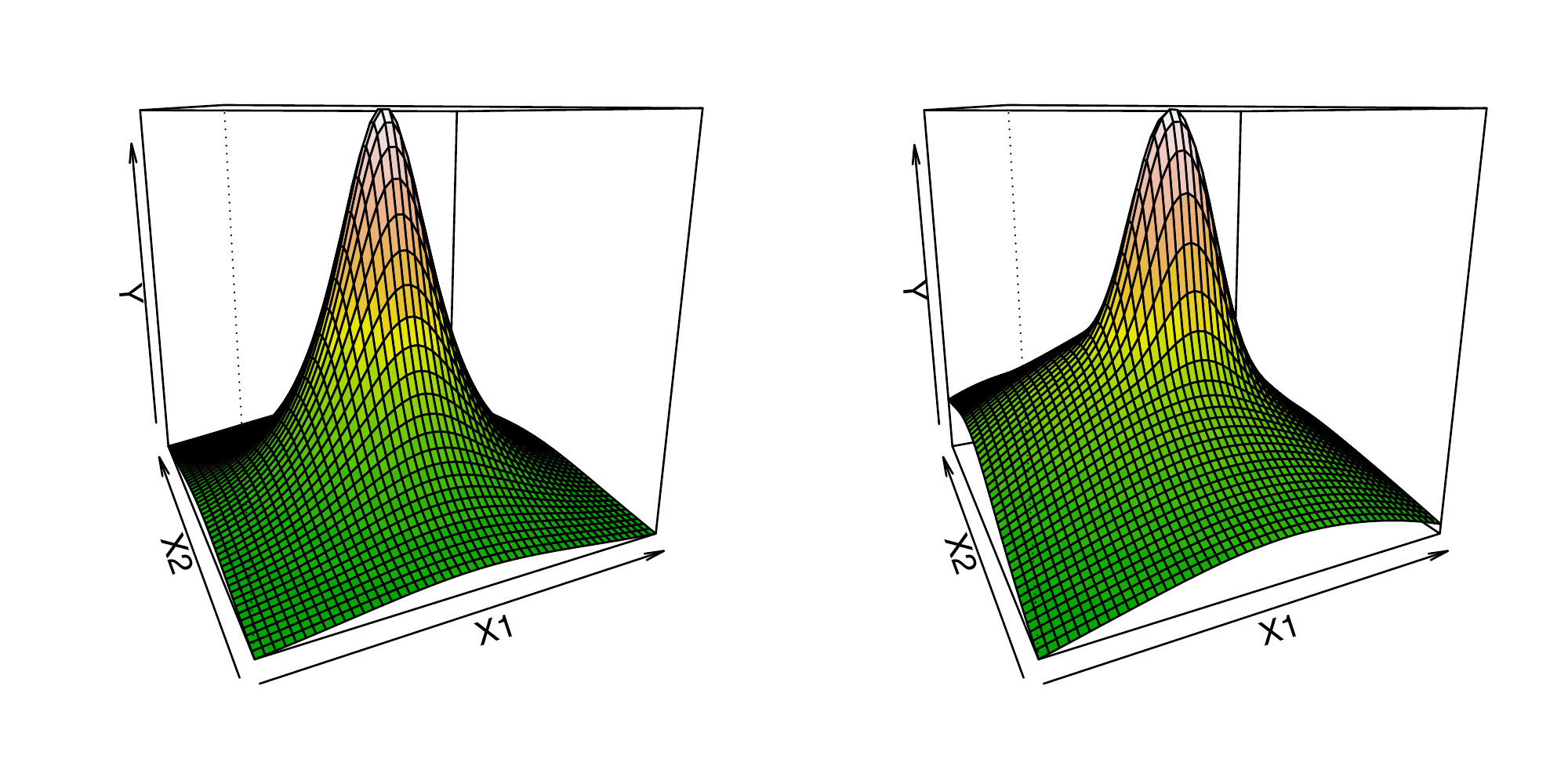}
\par\end{centering}
\caption{Results for the simulation Examples 3.1-3.3. Left panel: theoretical surface $\mathbb{E}\left[Y\mid\mathbf{X}\right]$;
Right panel: surface $\hat{\mathbb{E}}\left[Y\mid\mathbf{X}\right]$  predicted by S-SODA. \label{fig:E3}}
\end{figure}

We consider three examples to test the performance of using  formula (\ref{eq:Y_pred}) to predict $Y$, with $p=1000$ predictors simulated in the same way as scenario (a) in the previous section.
In order to visualize $\mathbb{E}\left[Y\mid\mathbf{X}_{\mathcal{P}}\right]$
and $\hat{\mathbb{E}}\left[Y\mid\mathbf{X}_{\tilde{\mathcal{P}}}\right]$
surface in a three-dimensional plot, we only had 2 relevant predictors, i.e., $\mathbf{X}_{\mathcal{P}}=\left(X_{1},X_{2}\right)$:
\begin{eqnarray*}
\text{Example 3.1:}\qquad Y & = & X_{1}+X_{2}+\sigma\epsilon,\\
\text{Example 3.2:}\qquad Y & = & X_{1}/\exp\left(X_{2}\right)+\sigma\epsilon,\\
\text{Example 3.3:}\qquad Y & = & \left(1+X_{1}^{2}+X_{2}^{2}\right)^{-1}+\sigma\epsilon,
\end{eqnarray*}
where $\sigma=0.2$ and $\epsilon\sim N\left(0,1\right)$. For each
example we simulated $n=500$ samples, and applied S-SODA to the simulated
data. S-SODA correctly identified $\tilde{\mathcal{P}}=\left\{ 1,2\right\} $.
We further used formula (\ref{eq:Y_pred}) with  $\hat{\boldsymbol{\mu}}$
and $\hat{\boldsymbol{\Sigma}}$ being the MLEs to predict $\hat{\mathbb{E}}\left[Y\mid\mathbf{X}_{\tilde{\mathcal{P}}}\right]$
with $H=25$ slices for each example. The results are 
shown in Figure~\ref{fig:E3}.
Encouragingly, it is observed that even though we do not know the true
functional form of $\mathbb{E}\left[Y\mid\mathbf{X}_{\mathcal{P}}\right]$,
our prediction $\hat{\mathbb{E}}\left[Y\mid\mathbf{X}_{\tilde{\mathcal{P}}}\right]$
well captures the landscape of $\mathbb{E}\left[Y\mid\mathbf{X}_{\mathcal{P}}\right]$
 in these examples.

\section{Real Data Applications \label{sec:realdata}}

We applied SODA, Lasso-Logistic, MDR and IIS-SQDA on real datasets
to compare their performances. We did not include the ZW method due to
its similarity with MDR. The classification accuracy of the selected models
were evaluated by 10-fold cross-validation after the variable selection.
For Lasso-Logistic and SODA, we used $\text{EBIC}_{0.5}$ as model
selection criterion. We consider three datasets: (1) a Michigan lung
cancer dataset analyzed in \cite{efron2009empirical} with large $p>5000$;
(2) the Ionosphere dataset, with $p=33$; and (3) the dataset Pumadyn, with $p=32$ and a continuous response.
The Ionosphere dataset was downloaded from UCI Machine Learning Repository\footnote{\texttt{https://archive.ics.uci.edu/ml/datasets/Concrete+Compressive+Strength}},
and the Pumadyn dataset was downloaded from DELVE (Data for Evaluating
Learning in Valid Experiments)\footnote{\texttt{http://www.cs.toronto.edu/\textasciitilde{}delve/data/pumadyn/desc.html}}.

\subsection{Michigan lung cancer dataset\label{sub:mich_lung}}
This dataset was published in \cite{beer2002gene},
in which researchers measured mRNA expression levels of $p=5,217$ genes
in  tumor tissues of 86 lung cancer patients. Among the 86 patients, 62 are labeled as in ``good status'', and 26 in ``bad status''. The goal is to classify new patients into one of two statuses. 
Results on this dataset are summarized in Table \ref{tab:mich_lung}.
IIS-SQDA  did not finish in 48 hours for this dataset, so we omitted its result. 

In the solution path of Lasso-Logistic, the lowest $\text{EBIC}_{0.5}$
was achieved at $112.2$ with $1$ gene, and the corresponding CV
error rate was $29\%$. SODA selected $2$ main effects and $2$ interaction
effects with the $\text{EBIC}_{0.5}$ score at $69.8$ and the CV
error rate at $11\%$. Similar to the prostate cancer dataset (see Supplemental Materials), SODA worked much better than Lasso-Logistic for finding the minimum
of $\text{EBIC}_{0.5}$ ($69.8$ vs $112.2$). Comparing results of
Lasso-Logistic and SODA selected models, it is obvious that interaction
effects selected by SODA contribute substantially to the classification
accuracy.

MDR failed to converge on this dataset. MDR selected as many
genes as possible until the number of selected genes was the same as the number of samples in the smaller class (26) and achieved a CV
error rate of $28\%$. The observation that MDR failed to converge for
both of these two large $p$ datasets illustrates the fact that the
QDA variable selection methods with joint normality assumption work
poorly for high-dimensional real datasets.

\begin{table}[h]
\begin{centering}
{\footnotesize{}}%
\begin{tabular}{|cc|cc|ccc|ccc|ccc|}
\hline 
\multicolumn{2}{|c|}{\textbf{\footnotesize{}Shrunken centroid }} & \multicolumn{2}{c|}{\textbf{\footnotesize{}Empirical Bayes}} & \multicolumn{3}{c|}{\textbf{\footnotesize{}MDR}} & \multicolumn{3}{c|}{\textbf{\footnotesize{}Lasso-Logistic}} & \multicolumn{3}{c|}{\textbf{\footnotesize{}SODA}}\tabularnewline
{\footnotesize{}\#P} & {\footnotesize{}CVE} & {\footnotesize{}\#P} & {\footnotesize{}CVE} & {\footnotesize{}$\Delta\text{BIC}$} & {\footnotesize{}\#P} & {\footnotesize{}CVE} & {\footnotesize{}$\text{EBIC}_{0.5}$} & {\footnotesize{}\#P} & {\footnotesize{}CVE} & {\footnotesize{}$\text{EBIC}_{0.5}$} & {\footnotesize{}\#M/\#I} & {\footnotesize{}CVE}\tabularnewline
\hline 
\hline 
{\footnotesize{}0} & {\footnotesize{}0.28} & {\footnotesize{}5} & {\footnotesize{}0.41} & {\footnotesize{}-353} & {\footnotesize{}1} & {\footnotesize{}0.27} & \textbf{\footnotesize{}112.2} & \textbf{\footnotesize{}1} & \textbf{\footnotesize{}0.29} & {\footnotesize{}111.2} & {\footnotesize{}1 / 0} & {\footnotesize{}0.33}\tabularnewline
{\footnotesize{}5} & {\footnotesize{}0.28} & {\footnotesize{}20} & {\footnotesize{}0.43} & {\footnotesize{}-177} & {\footnotesize{}2} & {\footnotesize{}0.26} & {\footnotesize{}113.9} & {\footnotesize{}2} & {\footnotesize{}0.25} & {\footnotesize{}104.1} & {\footnotesize{}2 / 0} & {\footnotesize{}0.25}\tabularnewline
{\footnotesize{}11} & {\footnotesize{}0.29} & {\footnotesize{}40} & {\footnotesize{}0.39} & {\footnotesize{}-178} & {\footnotesize{}3} & {\footnotesize{}0.25} & {\footnotesize{}122.2} & {\footnotesize{}3} & {\footnotesize{}0.25} & {\footnotesize{}98.2} & {\footnotesize{}3 / 0} & {\footnotesize{}0.21}\tabularnewline
{\footnotesize{}21} & {\footnotesize{}0.28} & {\footnotesize{}60} & {\footnotesize{}0.41} & {\footnotesize{}-165} & {\footnotesize{}4} & {\footnotesize{}0.24} & {\footnotesize{}121.0} & {\footnotesize{}4} & {\footnotesize{}0.20} & {\footnotesize{}89.6} & {\footnotesize{}4 / 0} & {\footnotesize{}0.17}\tabularnewline
{\footnotesize{}55} & {\footnotesize{}0.35} & {\footnotesize{}80} & {\footnotesize{}0.40} & {\footnotesize{}-156} & {\footnotesize{}5} & {\footnotesize{}0.25} & {\footnotesize{}131.5} & {\footnotesize{}5} & {\footnotesize{}0.22} & {\footnotesize{}79.2} & {\footnotesize{}5 / 0} & {\footnotesize{}0.12}\tabularnewline
{\footnotesize{}109} & {\footnotesize{}0.35} & {\footnotesize{}100} & {\footnotesize{}0.39} & {\footnotesize{}-134} & {\footnotesize{}8} & {\footnotesize{}0.24} & {\footnotesize{}144.5} & {\footnotesize{}6} & {\footnotesize{}0.23} & {\footnotesize{}94.4} & {\footnotesize{}5 / 1} & {\footnotesize{}0.12}\tabularnewline
{\footnotesize{}260} & {\footnotesize{}0.37} & {\footnotesize{}120} & {\footnotesize{}0.40} & {\footnotesize{}-132} & {\footnotesize{}11} & {\footnotesize{}0.29} & {\footnotesize{}150.5} & {\footnotesize{}7} & {\footnotesize{}0.25} & {\footnotesize{}$\vdots$} & {\footnotesize{}$\vdots$} & {\footnotesize{}$\vdots$}\tabularnewline
{\footnotesize{}567} & {\footnotesize{}0.38} & {\footnotesize{}140} & {\footnotesize{}0.40} & {\footnotesize{}-131} & {\footnotesize{}14} & {\footnotesize{}0.28} & {\footnotesize{}158.4} & {\footnotesize{}8} & {\footnotesize{}0.22} & \textbf{\footnotesize{}69.8} & \textbf{\footnotesize{}2 / 2} & \textbf{\footnotesize{}0.11}\tabularnewline
{\footnotesize{}1,173} & {\footnotesize{}0.40} & {\footnotesize{}160} & {\footnotesize{}0.42} & {\footnotesize{}-143} & {\footnotesize{}17} & {\footnotesize{}0.30} & {\footnotesize{}171.1} & {\footnotesize{}9} & {\footnotesize{}0.23} &  &  & \tabularnewline
{\footnotesize{}2,532} & {\footnotesize{}0.38} & {\footnotesize{}180} & {\footnotesize{}0.38} & {\footnotesize{}-146} & {\footnotesize{}20} & {\footnotesize{}0.27} & {\footnotesize{}177.6} & {\footnotesize{}10} & {\footnotesize{}0.23} &  &  & \tabularnewline
{\footnotesize{}5,217} & {\footnotesize{}0.38} & {\footnotesize{}200} & {\footnotesize{}0.40} & \textbf{\footnotesize{}-151} & \textbf{\footnotesize{}25} & \textbf{\footnotesize{}0.28} & {\footnotesize{}188.6} & {\footnotesize{}11} & {\footnotesize{}0.23} &  &  & \tabularnewline
\hline 
\end{tabular}
\par\end{centering}{\footnotesize \par}

\caption{Analysis results of the Michigan lung cancer dataset by five
methods. For Lasso-Logistic, MDR and SODA, the selected set with the lowest
BIC score is highlighted in bold font. $\Delta\text{BIC}$: For MDR
method, the difference of $\text{BIC}_{G}$ between two adjacent steps.
CVE: prediction error rate estimated by 10-fold cross-validation. \#P:
number of selected predictors. \#M / \#I: number of selected main
effect and interaction terms by SODA. \label{tab:mich_lung}}
\end{table}

\subsection{Ionosphere dataset}

This dataset is a two-class classification problem with 351 samples
and 32 predictors. Targets are ``Good'' and ``Bad'' radar returns
from the ionosphere. ``Good'' radar returns are those showing evidence
of some type of structure in the ionosphere, while ``Bad'' returns
do not. 

We applied Lasso-Logistic, MDR, IIS-SQDA and SODA to this dataset.
Since  the number of candidate predictors is not large, we  also ran Lasso-Logistic with all main effect terms and $32\times\left(32+1\right)/2=528$ interaction terms, which is referred to as Lasso-Logistic-2. Results are summarized in Table \ref{tab:Ionosphere}. In the solution path
of Lasso-Logistic, the lowest $\text{EBIC}_{0.5}$ was achieved at
$302.9$ with $6$ predictors, and the corresponding CV
error rate was $14\%$. Lasso-Logistic-2 selected 2 main effect terms
and 5 interaction terms with $\text{EBIC}_{0.5}$=$248.7$ and CV
error rate $8\%$. SODA selected $4$ main effect and $4$ interaction
effect terms with $\text{EBIC}_{0.5}$=$204.2$ and CV
error rate $6\%$. Again, SODA found a smaller $\text{EBIC}_{0.5}$ value than both Lasso methods. 

MDR method selected all 32 predictors and achieves CV
error rate $28\%$. IIS-SQDA selected 10 main effect and 96 interaction
terms and achieved CV error rate $16\%$. Since MDR
selected all 32 predictors, by definition MDR selected model is the
full QDA model. Comparing  this full QDA model with the SODA
selected model, we see that EBIC-based variable selection resulted in a much more interpretable model with a substantially
reduced classification error rate. 

\setlength\tabcolsep{4.5pt}

\begin{table}[h]
\begin{centering}
{\footnotesize{}}%
\begin{tabular}{|ccc|ccc|ccc|cc|ccc|}
\hline 
\multicolumn{3}{|c|}{\textbf{\footnotesize{}MDR}} & \multicolumn{3}{c|}{\textbf{\footnotesize{}Lasso-Logistic}} & \multicolumn{3}{c|}{\textbf{\footnotesize{}Lasso-Logistic-2}} & \multicolumn{2}{c|}{\textbf{\footnotesize{}IIS-SQDA}} & \multicolumn{3}{c|}{\textbf{\footnotesize{}SODA}}\tabularnewline
{\footnotesize{}$\Delta\text{BIC}$} & {\footnotesize{}\#P} & {\footnotesize{}CVE} & {\footnotesize{}$\text{EBIC}_{0.5}$} & {\footnotesize{}\#P} & {\footnotesize{}CVE} & {\footnotesize{}$\text{EBIC}_{0.5}$} & {\footnotesize{}\#M / \#I} & {\footnotesize{}CVE} & {\footnotesize{}\#M / \#I} & {\footnotesize{}CVE} & {\footnotesize{}$\text{EBIC}_{0.5}$} & {\footnotesize{}\#M / \#I} & {\footnotesize{}CVE}\tabularnewline
\hline 
\hline 
{\footnotesize{}-326} & {\footnotesize{}1} & {\footnotesize{}0.20} & {\footnotesize{}343.5} & {\footnotesize{}2} & {\footnotesize{}0.19} & {\footnotesize{}279.0} & {\footnotesize{}1 / 2} & {\footnotesize{}0.13} & \textbf{\footnotesize{}10 / 96} & \textbf{\footnotesize{}0.16} & {\footnotesize{}371.2} & {\footnotesize{}1 / 0} & {\footnotesize{}0.21}\tabularnewline
{\footnotesize{}-221} & {\footnotesize{}3} & {\footnotesize{}0.26} & {\footnotesize{}329.9} & {\footnotesize{}4} & {\footnotesize{}0.16} & {\footnotesize{}253.8} & {\footnotesize{}2 / 3} & {\footnotesize{}0.10} &  &  & {\footnotesize{}343.5} & {\footnotesize{}2 / 0} & {\footnotesize{}0.19}\tabularnewline
{\footnotesize{}-338} & {\footnotesize{}5} & {\footnotesize{}0.25} & \textbf{\footnotesize{}302.9} & \textbf{\footnotesize{}6} & \textbf{\footnotesize{}0.14} & {\footnotesize{}252.7} & {\footnotesize{}2 / 4} & {\footnotesize{}0.09} &  &  & {\footnotesize{}319.6} & {\footnotesize{}3 / 0} & {\footnotesize{}0.19}\tabularnewline
{\footnotesize{}-298} & {\footnotesize{}7} & {\footnotesize{}0.24} & {\footnotesize{}313.5} & {\footnotesize{}8} & {\footnotesize{}0.16} & \textbf{\footnotesize{}248.7} & \textbf{\footnotesize{}2 / 5} & \textbf{\footnotesize{}0.08} &  &  & {\footnotesize{}298.8} & {\footnotesize{}4 / 0} & {\footnotesize{}0.15}\tabularnewline
{\footnotesize{}-242} & {\footnotesize{}9} & {\footnotesize{}0.25} & {\footnotesize{}312.5} & {\footnotesize{}10} & {\footnotesize{}0.15} & {\footnotesize{}254.4} & {\footnotesize{}2 / 6} & {\footnotesize{}0.08} &  &  & {\footnotesize{}296.1} & {\footnotesize{}5 / 0} & {\footnotesize{}0.14}\tabularnewline
{\footnotesize{}-200} & {\footnotesize{}11} & {\footnotesize{}0.24} & {\footnotesize{}321.7} & {\footnotesize{}12} & {\footnotesize{}0.15} & {\footnotesize{}258.6} & {\footnotesize{}2 / 7} & {\footnotesize{}0.08} &  &  & {\footnotesize{}232.2} & {\footnotesize{}5 / 1} & {\footnotesize{}0.08}\tabularnewline
{\footnotesize{}-278} & {\footnotesize{}15} & {\footnotesize{}0.29} & {\footnotesize{}345.3} & {\footnotesize{}15} & {\footnotesize{}0.15} & {\footnotesize{}267.3} & {\footnotesize{}2 / 8} & {\footnotesize{}0.08} &  &  & {\footnotesize{}224.1} & {\footnotesize{}5 / 3} & {\footnotesize{}0.07}\tabularnewline
{\footnotesize{}-361} & {\footnotesize{}20} & {\footnotesize{}0.28} & {\footnotesize{}363.8} & {\footnotesize{}18} & {\footnotesize{}0.15} & {\footnotesize{}286.3} & {\footnotesize{}2 / 10} & {\footnotesize{}0.08} &  &  & {\scriptsize{}$\vdots$} & {\scriptsize{}$\vdots$} & {\scriptsize{}$\vdots$}\tabularnewline
{\footnotesize{}-434} & {\footnotesize{}25} & {\footnotesize{}0.30} & {\footnotesize{}383.1} & {\footnotesize{}22} & {\footnotesize{}0.15} & {\footnotesize{}290.3} & {\footnotesize{}2 / 11} & {\footnotesize{}0.08} &  &  & \textbf{\footnotesize{}204.2} & \textbf{\footnotesize{}4 / 4} & \textbf{\footnotesize{}0.06}\tabularnewline
\textbf{\footnotesize{}-130} & \textbf{\footnotesize{}32} & \textbf{\footnotesize{}0.28} & {\footnotesize{}445.4} & {\footnotesize{}30} & {\footnotesize{}0.16} & {\footnotesize{}312.0} & {\footnotesize{}2 / 13} & {\footnotesize{}0.08} &  &  &  &  & \tabularnewline
\hline 
\end{tabular}
\par\end{centering}{\footnotesize \par}

\caption{The summary of results on the Ionosphere dataset by the five methods.
$\Delta\text{BIC}$: For MDR method, the difference of $\text{BIC}_{G}$
between two adjacent steps. CVE: prediction error rate estimated by 10-fold
cross-validation. \#P: number of selected predictors. \#M / \#I: number
of selected main effect and interaction terms by SODA. \label{tab:Ionosphere}}
\end{table}

\subsection{Pumadyn dataset}

This dataset was synthesized from a realistic
simulation of the dynamics of a robotic arm. It has $n=8192$ samples, $p=32$ predictors, and a continuous response. The predictor set
includes angular positions, velocities and torques of the robot arm. 
The goal is to predict the angular acceleration of the robot arm's links. 
The samples are split into 4500 in-samples for modeling training, and 3692 out-samples for model evaluation.

\begin{figure}
\begin{centering}
\includegraphics[viewport=0bp 35bp 576bp 290bp,scale=0.75]{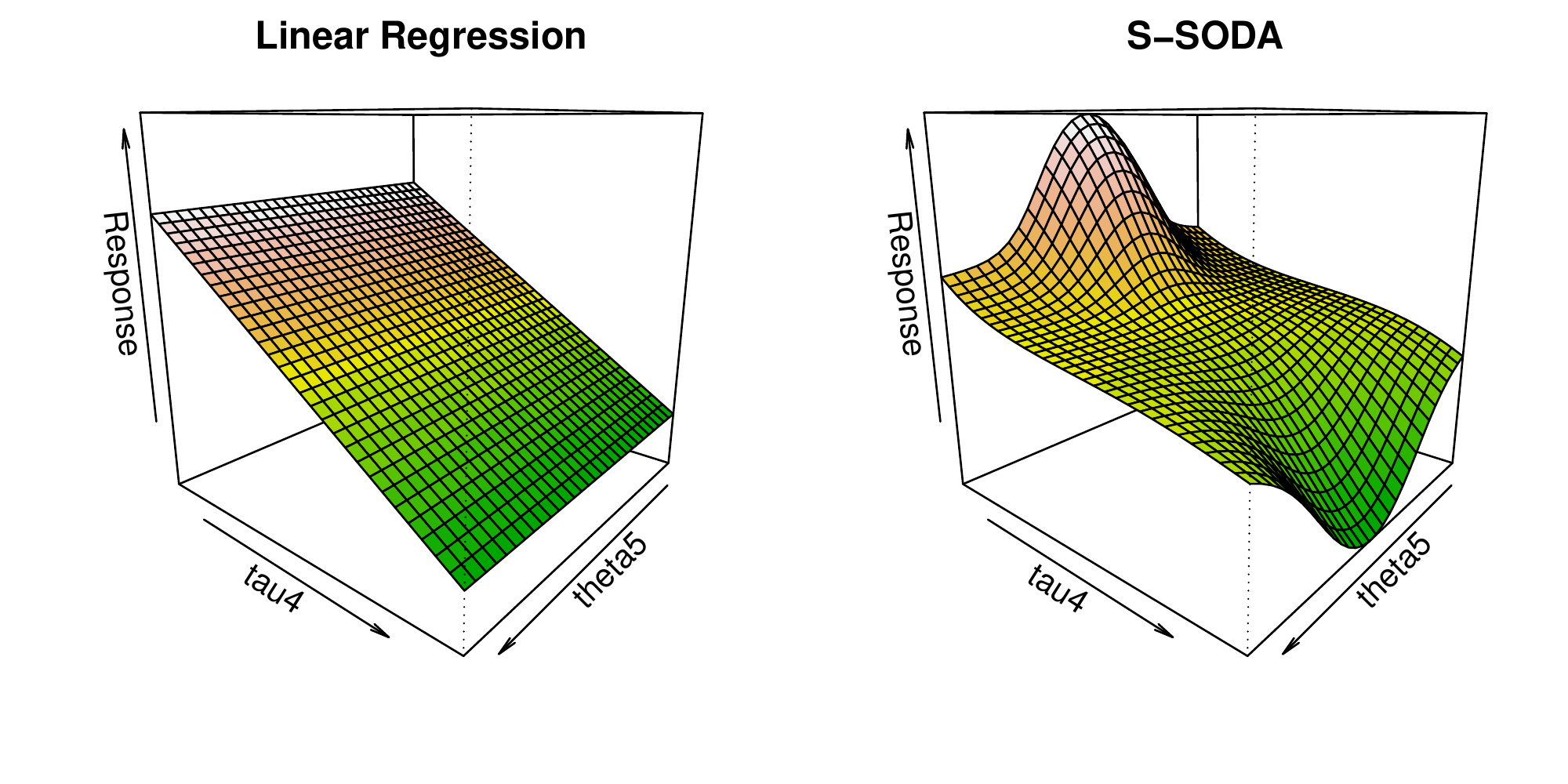}
\par\end{centering}

\caption{The predicted $\hat{\mathbb{E}}\left[Y\mid\mathbf{X}_{\left(\text{tau4},\text{theta5}\right)}\right]$
surface from linear model (left) and S-SODA (right).\label{fig:pumadyn_pred}}
\end{figure}

We trained the S-SODA model with $H=20$ for this dataset, and made the predictions
using formula (\ref{eq:Y_pred}). We also applied linear
regression with Lasso selection with/without interaction terms, denoted
as Lasso-Linear and Lasso-Linear-2, respectively.   The results are summarized in Table
\ref{tab:Pumadyn}. 
Lasso's highest out-sample correlation $r=0.477$ were achieved when selecting only 1 predictor
(named tau4). S-SODA selected two predictors (tau4 and theta5) and
achieved an out-sample correlation $r=0.707$. The predicted surfaces of $\hat{\mathbb{E}}\left[Y\mid\mathbf{X}_{\left(\text{tau4},\text{theta5}\right)}\right]$
 from the linear model and S-SODA, respectively, are shown in Figure \ref{fig:pumadyn_pred}.
The interaction between predictors tau4 and theta5 are captured by
S-SODA but not the linear model. From Table \ref{tab:Pumadyn} we can
also see that the interaction between tau4 and theta5 cannot be simply
captured by the multiplication term $X_{\text{tau4}}\cdot X_{\text{theta5}}$. 

\begin{table}[h]
\begin{centering}
{\footnotesize{}}%
\begin{tabular}{|cc|cc|cc|}
\hline 
\multicolumn{2}{|c|}{\textbf{\footnotesize{}Lasso-Linear}} & \multicolumn{2}{c|}{\textbf{\footnotesize{}Lasso-Linear-2}} & \multicolumn{2}{c|}{\textbf{\footnotesize{}S-SODA}}\tabularnewline
{\footnotesize{}\# Predictors} & {\footnotesize{}Out-$r$} & {\footnotesize{}\# M / \#I} & {\footnotesize{}Out-$r$} & {\footnotesize{}\# Predictors} & {\footnotesize{}Out-$r$}\tabularnewline
\hline 
\hline 
{\footnotesize{}1} & {\footnotesize{}0.477} & {\footnotesize{}1 / 0} & {\footnotesize{}0.477} & {\footnotesize{}1} & {\footnotesize{}0.469}\tabularnewline
{\footnotesize{}2} & {\footnotesize{}0.477} & {\footnotesize{}1 / 1} & {\footnotesize{}0.476} & {\footnotesize{}2} & {\footnotesize{}0.707}\tabularnewline
{\footnotesize{}3} & {\footnotesize{}0.476} & {\footnotesize{}1 / 2} & {\footnotesize{}0.474} &  & \tabularnewline
{\footnotesize{}4} & {\footnotesize{}0.476} & {\footnotesize{}1 / 3} & {\footnotesize{}0.473} &  & \tabularnewline
{\footnotesize{}5} & {\footnotesize{}0.476} & {\footnotesize{}1 / 4} & {\footnotesize{}0.473} &  & \tabularnewline
{\footnotesize{}10} & {\footnotesize{}0.474} & {\footnotesize{}1 / 10} & {\footnotesize{}0.469} &  & \tabularnewline
{\footnotesize{}20} & {\footnotesize{}0.472} & {\footnotesize{}1 / 20} & {\footnotesize{}0.464} &  & \tabularnewline
{\footnotesize{}30} & {\footnotesize{}0.472} & {\footnotesize{}1 / 30} & {\footnotesize{}0.459} &  & \tabularnewline
\hline 
\end{tabular}
\par\end{centering}{\footnotesize \par}

\caption{Analysis  results of the Pumadyn dataset by the three methods.
\#P: the number of selected predictors. \#M / \#I: the number of selected
main effect and interaction terms by Lasso on linear model with interaction
terms. Out-$r$: the out-sample correlation $r$. \label{tab:Pumadyn}}
\end{table}

\section{Concluding Remarks}

A somewhat striking observation in this article is that the proposed stepwise selection algorithm SODA, which is guided by EBIC and based on the classic stepwise regression idea with a twist for efficiently searching for interaction terms, out-performed all known advanced methods, such as those based on $L_1$ regularizations, in terms of  variable selection accuracy, prediction accuracy, and robustness in a variety of settings when the joint distribution of the predictors do not "behave nicely." 
In contrast to \cite{murphy2010variable}, \cite{zhang2011bic}, and \cite{maugis2011variable},
the consistency of SODA does not require the joint normality assumption
of relevant and irrelevant predictors. Compared to IIS in \cite{fan2015innovated},
SODA's forward variable addition does not need the normal assumption and
does not need to estimate large precision matrices. 

It is worth noting that even for logistic regression models with only main effects,  we consistently observed that SODA performed better than or similarly to  Lasso-logistic in terms of both the $\text{EBIC}_{0.5}$ score and the CV error rate under various settings, especially  when the predictors are highly correlated or the joint distribution of the predictors is long-tailed. In Supplemental Materials,  
we also observed that Lasso-logistic failed miserably when the `incoherence condition''  \citep{ravikumar2010high} was violated in linear logistic models, whereas SODA still performed robustly.
These indicate that EBIC is a good criterion to follow and SODA is a better optimizer of EBIC than Lasso.  Indeed, when one moves away from the $L_1$ regularization realm but adopts the $L_0$ regularization framework (such as AIC, BIC, EBIC), Lasso can no longer guarantee to find the optimal solution and is more {\it ad hoc} than stepwise approaches. 

LDA and QDA complement each other in terms of the bias-variance trade-off.
Given finite observations, LDA is simpler and more robust when the
response $Y$ can be explained well by the linear effects of $\mathbf{X}$.
QDA has the ability to exploit interaction effects, which may contribute
dramatically to the classification accuracy, but also has many more parameters
to estimate and is more vulnerable to including noise predictors. SODA
is designed to be adaptive in the sense that it automatically chooses
between LDA and QDA models and takes advantage of both sides. Instead
of selecting predictors, SODA selects individual main and interaction
terms, which enables SODA to simultaneously utilize interaction
terms and avoid including a large number of unnecessary terms.

An interesting and also somewhat surprising twist of SODA is its extension S-SODA for dealing with the variable selection problem for semi-parametric models with continuous responses. Our simulation results demonstrated that the simple idea of slicing (aka {\it discretizing}) the response variable can bring a lot to the table, especially coupled with stepwise variable selection tools such as N-SIRI \citep{jiang2014variable} and S-SODA. Even for linear models, S-SODA performed competitively with Lasso, and outperformed other linear or near-linear methods such as  hierNet and DC-SCAD  when the joint distribution of the covariates is long-tailed.  Compared with existing SIR-based methods, SODA does not require the linearity and constant variance conditions and enjoys a much improved robustness.

A main limitation of SODA is that the stepwise detectable condition might not hold when main effects are very weak or nonexistent but the interaction effects are strong.  This is a generic issue that troubles almost all methods unless we put all interaction terms into the set of candidate variables subject to selection.
In empirical studies we found that SODA worked well and had better performances
compared to other methods for both simulated and real-data examples, which suggests that this may not be a serious issue in many real applications. 
Indeed, even for QDA models it is quite unusual and nearly pathological to construct mean vectors and covariance matrices that result in a discriminant function with no main effects but only interaction terms.

The Implementation of SODA and S-SODA procedures is available in the
R package \texttt{sodavis} on CRAN (\texttt{http://cran.us.r-project.org}).

\bibliography{refs}
\bibliographystyle{chicago}

\end{document}